\documentclass[twocolumn, twocolappendix]{aastex631}

\usepackage{amsmath}

\begin{document}

\title{L1448 IRS3B: Dust Polarization Aligned with Spiral Features, Tracing Gas Flows}

\shorttitle{L1448 IRS3 Polarization}
\shortauthors{Looney et al.}

\correspondingauthor{Leslie W. Looney}
\email{lwl@illinois.edu}

\author[0000-0002-4540-6587]{Leslie W. Looney}
\affiliation{Department of Astronomy, University of Illinois, 1002 West Green St, Urbana, IL 61801, USA}
\affil{National Radio Astronomy Observatory, 520 Edgemont Rd., Charlottesville, VA 22903 USA} 

\author[0000-0001-7233-4171]{Zhe-Yu Daniel Lin}
\affiliation{University of Virginia, 530 McCormick Rd., Charlottesville, Virginia 22904, USA}
\affiliation{Earth and Planets Laboratory, Carnegie Institution for Science, 5241 Broad Branch Rd. NW, Washington, DC 20015, USA}

\author[0000-0002-7402-6487]{Zhi-Yun Li}
\affiliation{University of Virginia, 530 McCormick Rd., Charlottesville, Virginia 22904, USA}
\affiliation{Virginia Institute of Theoretical Astronomy, University of Virginia, Charlottesville, Virginia 22904, USA}

\author[0000-0002-6195-0152]{John J. Tobin}
\affil{National Radio Astronomy Observatory, 520 Edgemont Rd., Charlottesville, VA 22903 USA} 

\author{Martin Radecki}
\affiliation{Department of Astronomy, University of Illinois, 1002 West Green St, Urbana, IL 61801, USA}

\author[0000-0003-0002-3723]{Syzygy Butte}
\affiliation{Department of Astronomy, University of Illinois, 1002 West Green St, Urbana, IL 61801, USA}

\author{Ian W. Stephens}
\affiliation{Department of Earth, Environment, and Physics, Worcester State University, Worcester, MA 01602, USA}

\author[0000-0001-5811-0454]{Manuel Fern\'{a}ndez-L\'{o}pez}
\affiliation{Instituto Argentino de Radioastronom{\'i}a, CCT-La Plata (CONICET), C.C.5, 1894, Villa Elisa, Argentina}

\author[0000-0002-8537-6669]{Haifeng Yang}
\affiliation{Institute for Astronomy, School of Physics, Zhejiang University, Hangzhou, 310027 Zhejiang, China}

\author[0000-0002-9239-6422]{Nickalas K. Reynolds}
\affiliation{Homer L. Dodge Department of Physics and Astronomy, The University of Oklahoma, 440 W Brooks St, Norman, OK, 73019 USA}

\author[0000-0002-9209-8708]{Patrick Sheehan}\affil{National Radio Astronomy Observatory, 520 Edgemont Rd., Charlottesville, VA 22903 USA} 

\author[0000-0003-4022-4132]{Woojin Kwon}
\affiliation{Department of Earth Science Education, Seoul National University, 1 Gwanak-ro, Gwanak-gu, Seoul 08826, Republic of Korea}
\affiliation{SNU Astronomy Research Center, Seoul National University, 1 Gwanak-ro, Gwanak-gu, Seoul 08826, Republic of Korea}
\affiliation{The Center for Educational Research, Seoul National University, 1 Gwanak-ro, Gwanak-gu, Seoul 08826, Republic of Korea}

\author{Rachel Harrison}
\affiliation{School of Physics and Astronomy, Monash University, Clayton VIC 3800, Australia}

\author[0009-0008-9409-6686]{Allen North}
\affiliation{Department of Astronomy, University of Illinois, 1002 West Green St, Urbana, IL 61801, USA}



\begin{abstract}

Circumstellar disk dust polarization in the (sub)millimeter is, for the most part, not from dust grain alignment with magnetic fields but rather indicative of a combination of dust self-scattering with a yet unknown alignment mechanism that is consistent with mechanical alignment. 
While the observational evidence for scattering has been well established, that for mechanical alignment is less so. Circum-multiple dust structures in protostellar systems provide a unique environment to probe different polarization alignment mechanisms. We present ALMA Band 4 and Band 7 polarization observations toward the multiple young system L1448 IRS3B. The polarization in the two Bands is consistent with each other, presenting multiple polarization morphologies. On the size scale of the inner envelope surrounding the circum-multiple disk, the polarization is consistent with magnetic field dust grain alignment. On the very small scale of compact circumstellar regions, we see polarization that is consistent with scattering around source $a$ and $c$, which are likely the most optically thick components. Finally, we see polarization that is consistent with mechanical alignment of dust grains along the spiral dust structures, which would suggest that the dust is tracing the relative gas flow along the spiral arms. If the gas-flow dust grain alignment mechanism is dominant in these cases, disk dust polarization may provide a direct probe of the small-scale kinematics of the gas flow relative to the dust grains.

\end{abstract}



\section{Introduction} \label{sec:intro}

Dust continuum polarization was first unequivocally detected in a circumstellar disk a decade ago \citep{Stephens2014}.
Dust polarized emission is routinely used to infer the magnetic field morphology in star formation regions down to the scale of the inner envelopes of young protostars \citep[e.g.,][]{Hull2014,Cox2018,Galametz2018,LeGouellec2020,Maury2022,Huang2024}, but over the last few years it has become clear that the polarization in circumstellar disks has multiple mechanisms, including dust self-scattering \citep[e.g.,][]{Kataoka2015,Yang2016} and some yet unknown dust grain alignment mechanism \citep{Yang2019,Mori2021,Hoang2022,Tang2023,Lin2024a}. In fact, direct magnetic field alignment appears to be unlikely in the vast majority of circumstellar disk cases with the possible exceptions of some circumbinary disks, such as 
BHB 07-11 \citep{Alves2018}, VLA 1623 \citep{Sadavoy2018}, HD 142527 \citep{Ohashi2018,Ohashi2025}, and GG Tau \citep{Tang2023}, and the single system
TMC1A \citep{Aso2021}, given that toroidal magnetic field models generally reproduce the observed polarization for those sources. However, higher resolution observations are critical to model the polarization emission uniquely \citep[e.g.,][]{Dent2019,Lin2020,Stephens2023}.

Observed polarization due to dust self-scattering is possible if the dust grains are comparable in size to the observing wavelength, there is some optical depth, and there is some anisotropy of the radiation field \citep[e.g.,][]{Kataoka2015}. When observed in inclined circumstellar disks, the polarization morphology from scattering will be along the minor axis of the disk with fractional polarizations (polarized emission / Stokes I emission) of $\sim$1\% \citep{Kataoka2015,Yang2016}.


Elongated dust grains may be aligned in the disk, which can also produce polarized emission. The most commonly accepted alignment mechanism in the ISM is grain alignment from magnetic fields via Radiative Alignment Torques \citep[often called B-RAT, e.g.,][]{Dolginov1976,Draine1997}, which typically has fractional polarizations of $\sim$5-15\%. 
However, recent work has suggested that it is unlikely for $\gtrapprox$100 $\mu$m dust grains, as may be expected in disks, to be aligned by magnetic fields \citep{Yang2021}.
Dust grain alignment due to purely Radiative Alignment Torques is often called k-RAT with expected fractional polarizations $\sim$1\% \citep[e.g.,][]{Tazaki2017,Hoang2022}, but purely radiative alignment was shown to likely be a less important mechanism in circumstellar disks \citep{Yang2019,Mori2019}. 


In addition to magnetic and radiative torques, dust grains can be aligned by mechanical torques. For helical dust grains, 
if the Larmor precession time is shorter than the mechanical precession time, then the grains can still become aligned to the B-field even if the origin of the torques was completely from mechanical torques, often called B-MET \cite[e.g.][]{Hoang2016}.  If the Larmor precession time is longer than the mechanical precession time, then the grains can become aligned with their short axis to the gas flow, often called v-MET (or v-MAT) \citep[e.g.,][]{Lazarian2007,Hoang2018,Hoang2022}.
In some cases, the dust helicity allows the dust grains to be rotated up suprathermally with only subsonic drift \citep{Hoang2018}.
In addition, dust grains with or without helicity may be aligned by mechanical alignment with their long axis aligned  to the gas flow \citep{Gold1952}, although this is a stochastic torque and not as strongly driven as in the v-MET case.
On the other hand, recent work by \cite{Lin2024b} suggests that gas drag alone can align dust grains through relative gas-dust motion, without the supersonic requirement of \cite{Gold1952}, when the dust grain's center of mass is slightly offset from its geometric center, which they call badminton birdie alignment.
In other words, there are many possible dust grain alignment mechanisms that can cause dust polarization in disks, in addition to scattering, and observations are beginning to constrain the dominant polarization origins in disks. 

Circumstellar disks with multiwavelength observations have shown interesting variations in some disk polarization observations with the emission being consistent with dust self-scattering at the short wavelengths of ALMA (i.e., polarization angles aligned with the disk minor axis), smoothly transitioning to an azimuthal pattern 
at longer wavelengths \citep[e.g.,][]{Kataoka2017,Harrison2019,Ohashi2023,Lin2024a}.  
For example, in the cases of HL Tau and DG Tau, the disks have both a higher polarization fraction and a more azimuthal polarization orientation at the longer wavelength of Band 3 than at Band 7 \citep[e.g.,][]{Kataoka2017,Ohashi2023}, which is not possible from self-scattering polarization alone. Note that radiative grain alignment does not match the azimuthal pattern observations as k-RAT would produce a circular morphology (like a bulls-eye) whereas the observations at long wavelengths instead produce elliptical patterns \citep{Yang2019,Mori2019}. On the other hand, some disks show polarization that is dominated by dust self-scattering at both ALMA bands 3 and 7 \citep[e.g.,][]{Harrison2019,Harrison2024}.

Recent models have shown that the polarization across ALMA bands may be explained as an optical depth effect. At high optical depth, the polarization is dominated by scattering, while at lower optical depth, the polarization is dominated by aligned grains, with a smooth transition between the two limits \citep{Stephens2017,Mori2019,Lin2022,Lin2024a}. 
High resolution Band 7 observations of HL Tau with its gaps and rings have reinforced this viewpoint as the rings, which have higher optical depth, are dominated by dust self-scattering, while the gaps are dominated by aligned grains \citep{Stephens2023}.  However, the dust scattering of the HL Tau rings still places constraints on the maximum dust grains size (assuming a spheroidal, non-porous power-law dust grain distribution) of $\sim$100$\mu$m, which is hard to reconcile with the maximum dust grain size of a few millimeters from dust continuum emission modeling \citep{Kwon2011,Carrasco-Gonzalez2019}.

In order to explore the polarization of disks, we use the young multiple protostellar system L1448 IRS3, as circumbinary disks to date are among the few sources with posited combinations of magnetic field and scattering polarization mechanisms. 
The L1448 IRS3 system is at a distance of 288 pc based on recent constraints \citep{Ortiz2018,Zucker2019}. The system
consists of three distinct sources at $\sim$1\arcsec\ resolution \citep[e.g.,][]{Looney2000}, the brightest of these in the (sub)millimeter dust continuum is L1448 IRS3B, which is projected in the northern blushifted lobe of the outflow cavity of the source L1448-mm. In addition, the protostar L1448 IRS3A, which is $\sim$2300 au to the north, is interacting gravitationally with L1448 IRS3B \citep{Kwon2006,Nick2021,Reynolds2023,Gieser2024}.

L1448 IRS3B has long thought to be a young (likely $<$150 kyrs) Class 0 protostar \citep[e.g.,][]{Dunham2015,Podio2021,Reynolds2023}. 
This is consistent with the recently compiled SED from the eHOPS catalog\footnote{https://irsa.ipac.caltech.edu/data/Herschel/eHOPS\\ /ehops\_perseus\_seds\_protostars.html}, which gives a combined bolometeric temperature for L1448 IRS3A and B of 25.6 K \citep{Pokhrel2023}.  L1448 IRS3A is considered to be less embedded, perhaps being Class I \citep[e.g.,][]{Reynolds2023}. L1448 IRS3C (L1448 NW) is 17$\arcsec$ to the northwest of IRS3B and has a bolometeric temperature of 21.8 K \citep{Pokhrel2023}, which also suggests a Class 0 source \citep[also see,][]{Enoch2009}.
IRS3C is a close binary (separation 72~au) with the two protostars having their own disks \citep{Reynolds2023}.

The L1448 IRS3B protostellar multiple source has a circum-multiple disk caught in the act of fragmenting creating at least three and likely four protostars \citep{John2016,Nick2021,Reynolds2023}. 
There are three compact dust emission peaks detected in ALMA Bands 6 and 7 that are thought to be three protostellar sources. Source $c$, to the east of the central region, is the brightest in the dust continuum, but it is thought to be the youngest protostar with an upper limit on the protostellar mass of 0.2 M$_\odot$ whereas the combined mass of sources $a$ and $b$, near the center, is $\sim$1.2 M$_\odot$ \citep{Nick2021}. The circum-multiple disk has a Keplerian rotation pattern that places the kinematic center of the system near source $a$. Recent high resolution observations in Band 6 detect a fourth compact source (source $d$) near the kinematic center of the system surrounded by a ring of material, implying that this may be a quadruple protostellar system \citep{Reynolds2023}.
Note that these stellar masses are still comparable to other Class 0 source masses seen in multiple low-mass systems such as Ced110 IRS4A/B or  R CrA IRAS 32A/B \citep{Yen2024}.

Previous polarization observations of L1448 IRS3B \citep[][using the Berkeley-Illinois-Maryland Association array]{Kwon2006} showed $\lambda$ = 1.3~mm linear dust polarization on the scale of $\sim$5$\arcsec$ in the inner envelope that was $\sim$5\% polarized with a polarization angle of -70$^\circ$ (east of north), if only the central vectors are considered, which was also consistent with larger scale ($\sim$13$\arcsec$ resolution) James Clerk Maxwell Telescope (JCMT) SCUBAPOL $\lambda$ = 850 $\mu$m dust linear polarization observations \citep{Kwon2006}. More recent JCMT POL-2 observations with increased sensitivity are also consistent with the BIMA and SCUBAPOL observations (Kim et al., in prep). With $\lambda$ = 870 $\mu$m Submillimeter Array (SMA) dust observations at $\sim$3$\arcsec$ resolution, a dust polarization fraction of $\sim$3\% with a polarization angle of -67$^\circ$ (east of north) has been observed \citep{Galametz2018}. In all cases, the polarization is generally consistent with dust grains in the inner envelope of L1448 IRS3B aligned with an inferred large scale magnetic field at a PA of 23$^\circ$, although it has yet to be directly proven.

In this paper, we present ALMA Band 4 and 7 dust polarization observations of the L1448 IRS3B system. We use the dust polarization to infer possible polarization mechanisms in the circum-multiple disk and inner envelope of L1448 IRS3B, finding evidence of three polarization mechanisms.  In \S 2, we present the observations and data reduction, in \S 3, we present the results of the observations, in \S 4, we discuss the results and place constraints on the polarization mechanisms, in \S 5, we present our conclusions, and in the Appendix, we provide our observations of other sources in the field: images of L1448 IRS3A, L1448 IRS3C (or L1448 NW), and a newly detected point source in Band 4 that is likely a background galaxy.

\section{Observations and Data Reduction}

The Atacama Large Millimeter/submillimeter Array (ALMA) observations in full polarization at continuum Band 4 and Band 7 (i.e. $\lambda$ = 2.1 mm and 873 $\mu$m, respectively) were obtained as part of project 2021.1.00276.S. The observational details are described in Table \ref{tab:obs}. The observations were calibrated by the National Radio Astronomy Observatory staff (since polarization is a non-standard observing mode) within CASA, Common Astronomy Software Applications \citep{casa2022}, version 6.4.1.12. 

\begin{table*}[h]
    \centering
    \caption{Observation Information} \label{tab:obs}
\begin{tabular}{cc c c c c c c c c c c} 
\hline \hline
Band & Observation & Time on       & Conf.         & Ants.$^{a}$ & Baselines & MRS$^{b}$ & PWV$^{c}$   & \multicolumn{3}{c}{Calibrators}       \\ 
 & Date         & Source       &  &   &    &       &       & Bandpass      & Phase         & Pol. & Flux \\
& (y-m-d) & (h:mm.m)   &         &           & (m)   &    & (mm)  &               &               &    &   \\

\hline

4 & 2022-07-17  & 3:50.4    & C6 & 42        & 15 - 2617$^d$   & 1$\farcs$6    & 0.3 & J0237+2848    & J0336+3218    & J0348-2749 & J0237+2848\\
 
7&  2021-11-17  & 2:48.7    & C7   & 49        & 41 - 3638$^e$  & $2\farcs$6 & 20.6 & J0237+2848    & J0336+3218   & J0348-2749 & J0237+2848\\

\hline
\end{tabular}
\tablenotetext{}{$^{a}$Maximum number of antennas available during observation.$^{b}$ Maximum recoverable scale. $^{c}$ Average perceptible water vapor measurement. $^{d}$ 7.1 to 1,246 k$\lambda$. $^{e}$ 47 to 4,167 k$\lambda$. }
\end{table*}

\begin{table*}[h]
    \centering
    \caption{Map Details} \label{tab:maps}
\begin{tabular}{cc c c c c c c} 
\hline \hline
Band & Robust & Beam  & $\sigma_I^a$   &  $\sigma_Q$ &  $\sigma_U$ & $\sigma_P$ & \\
&    &   & ($\mu$Jy/bm)    &  ($\mu$Jy/beam)     &     ($\mu$Jy/bm)  & ($\mu$Jy/bm)   \\

\hline

4 & 0.5 & 0$\farcs$25 $\times$ 0$\farcs$18, 1.9$^{\circ}$ & 16.7 & 7.3 & 7.5 & 7.4\\

4 & 1.0 & 0$\farcs$30 $\times$ 0$\farcs$22, 0.2$^{\circ}$ & 26.7 & \nodata & \nodata & \nodata \\
 
7& 0.5 & 0$\farcs$15 $\times$ 0$\farcs$10, 11.7$^{\circ}$ & 114.0 & 25.0 & 26.0 & 25.5 \\

7$^b$ & 0.5 & 0$\farcs$25 $\times$ 0$\farcs$18, 1.9$^{\circ}$ & 307.1 & 34.2 & 38.6 & 36.5 \\

\hline
\end{tabular}
\tablenotetext{}{$^{a}$Note that $\sigma_I$ is typically larger than the other RMS due to dynamic range limitations. $^{b}$ Smoothed to match the Band 4 beam. }
\end{table*}

After the data were calibrated, we further performed self-calibration using the \texttt{auto-selfcal}\footnote{https://github.com/jjtobin/auto\_selfcal} script with \texttt{refantmode=`strict'}. 
The Band 4 observations were iteratively self-calibrated, phase only, with intervals of \texttt{inf-EB}, \texttt{inf}, 26.21s, 8.06s, 4.03s, and \texttt{int}. The \texttt{inf-EB} interval averages the 
entire execution block (when combined with \texttt{combine='scan,spw'} during the \texttt{gaincal} step of applying the derived gains to the visibilities) and corrects the time invariant phase errors in the data.
At the shorter time step, \texttt{int} does not do any time averaging and corrects time variant phase errors down to the 
integration time.  
Afterwards, a round of amplitude and phase self-calibration was performed with an interval of 300s to further improve the calibration.  
The Band 7 observations were self-calibrated, phase only, with intervals of \texttt{inf-EB}, \texttt{inf}, and finally 48.38s. 
Afterwards, a round of amplitude and phase self-calibration was performed with an interval of \texttt{inf} to further improve the calibration.  

The Stokes I, Q, and U maps were then created using the CASA task \texttt{tclean} with a \texttt{Briggs} weighting \texttt{Robust} weighting of 0.5, which is a good compromise between resolution and sensitivity \citep{Briggs1995}. The measured linear polarization intensity map is created using $P_m = \sqrt{Q^{2} + U^{2}}$.  
The linear polarized intensity noise is defined as $\sigma_{P} = \sqrt{(\sigma_{Q}^{2} + \sigma_{U}^{2})} / 2$, where $\sigma_{Q}$ and $\sigma_{U}$ are the noises in the Stokes Q and I maps given in Table \ref{tab:maps}. 
Following \cite{Wardle1974,Vaillancourt2006,Hull2015}, the linear polarization at low S/N ($<$~3$\sigma_P$) was debiased using the probability density function (PDF)

\begin{align} 
    \text{PDF}(P | P_{m}, \sigma_{P}) = 
    \frac{P}{ \sigma_{P}^{2} } I_{0} 
    \bigg(\frac{ P P_{m} }{ \sigma_{P}^{2} } \bigg) 
    \exp{ \bigg[ - \frac{ ( P_{m}^{2} + P^{2})}
    {2 \sigma_{P}^{2}} \bigg]},
\end{align}
which provides the true linear polarized intensity $P$ given our measured linear polarized intensity and the noise level $\sigma_{P}$, using a zeroth-order modified Bessel function of the first kind $I_{0}$.  For higher S/N, the approximation $P = \sqrt{ Q^{2} + U^{2} -  \sigma_{P}^{2}}$ was used to debias the linear polarization. The polarization angle is calculated using $\chi = \frac{1}{2} \arctan (\frac{U}{Q})$, where $\sigma_{\chi} = \frac{1}{2} \frac{ \sigma_{P} }{ P }$.

As suggested in the ALMA handbook, the absolute flux calibration accuracy is adopted to be $\sim$10\%. 

\section{Results}

\subsection{L1448 IRS3B Results}\label{IRS3B Results}

\begin{figure*}
  \centering
    \includegraphics[angle=0,width=0.90\textwidth]{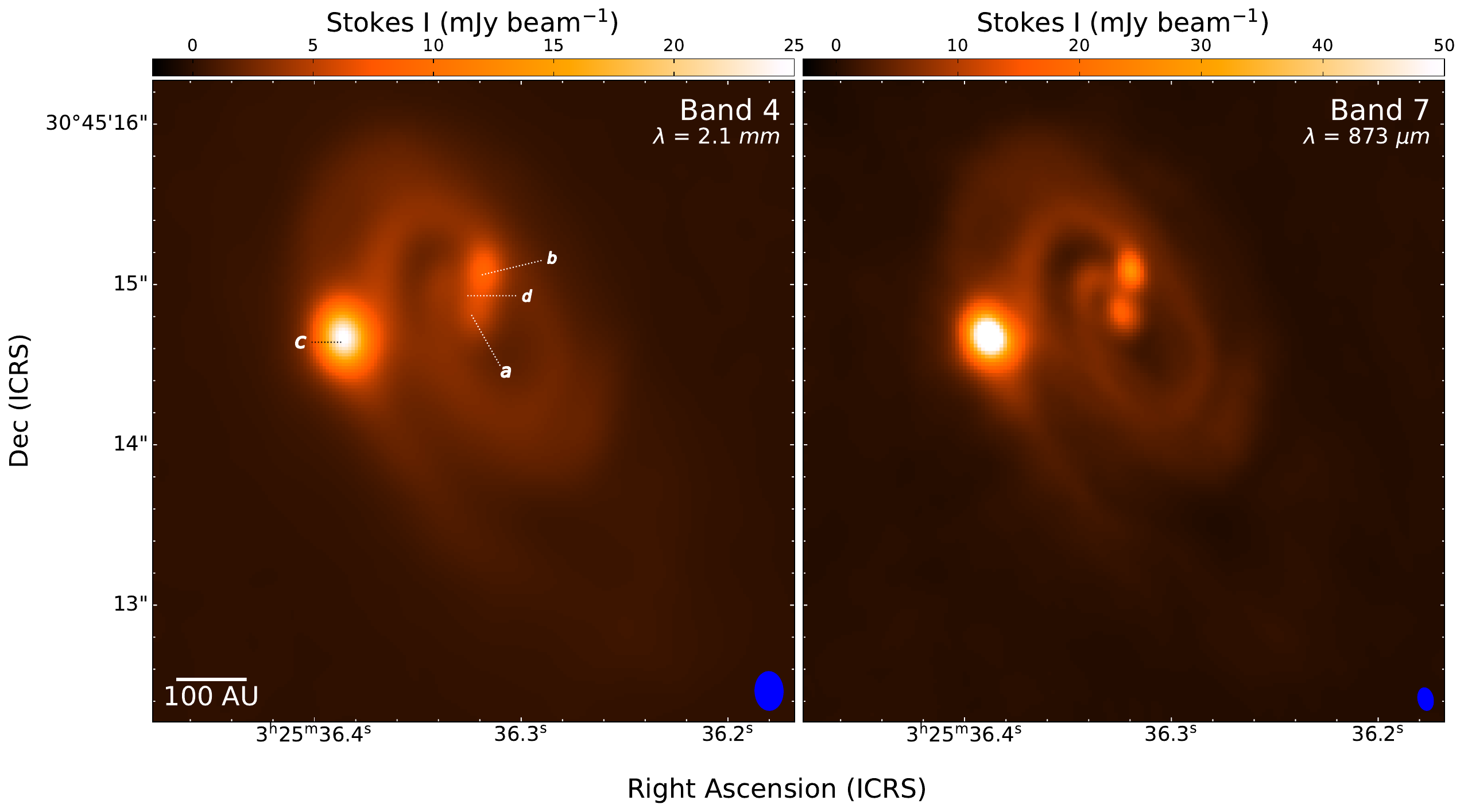}
\caption{ALMA Band 4 and Band 7 dust continuum maps of L1448 IRS3B. Source $a$, $b$, $c$, and $d$ are labeled in Band 4 \citep[from][]{Reynolds2023}. The beams, sizes listed in Table \ref{tab:maps}, are shown in blue in the bottom right corners.
 }
  \label{stokesI}
\end{figure*}

In Figure \ref{stokesI}, we present the dust continuum images of L1448 IRS3B in Band 4 and Band 7. These are the highest resolution continuum observations of the system at Band 4 to date. Band 4 is more sensitive to larger structures due to its shorter baselines (see Table \ref{tab:obs}). Nonetheless, the Band 4 map overall matches the continuum structures seen in Band 7. The Band 7 image is consistent with the slightly higher resolution dust continuum Band 7 observations in \cite{Nick2021}. We detect at both wavelengths the arcs or spiral structures and the three compact sources defined by \cite{John2016}, labeled $a$, $b$, and $c$ in Figure \ref{stokesI}.

The spiral structures are very clear in both Bands. However, the C$^{17}$O (3-2) velocity field measured at 0$\farcs$2 spatial resolution reveals Keplerian motion around a location near source $a$ without any clear perturbations caused by the spiral structures \citep{Nick2021}.
On the northwest of the disk, there is a sharp decrease in dust emission at both Bands that is nearly opposite of source $c$. 

Higher resolution ($\sim$8 au) Band 6 ALMA observations of IRS3B detected a fourth compact source, named source $d$, near the kinematic center of the system \citep{Reynolds2023}, also labeled in Figure \ref{stokesI}. They also resolved a ring (whose outer edges can be seen near the center of Band 7 in Figure \ref{stokesI}) surrounding source $d$ and enclosing source $a$; source $a$ is just inside the ring, and source $b$ is just outside the ring. In fact in the Band 7 map, source $a$ is slightly extended to the northeast, which is perhaps a hint of source $d$ or more likely a blending of the ring and source $a$.


\begin{figure*}
  \centering
    \includegraphics[angle=0,width=0.90\textwidth]{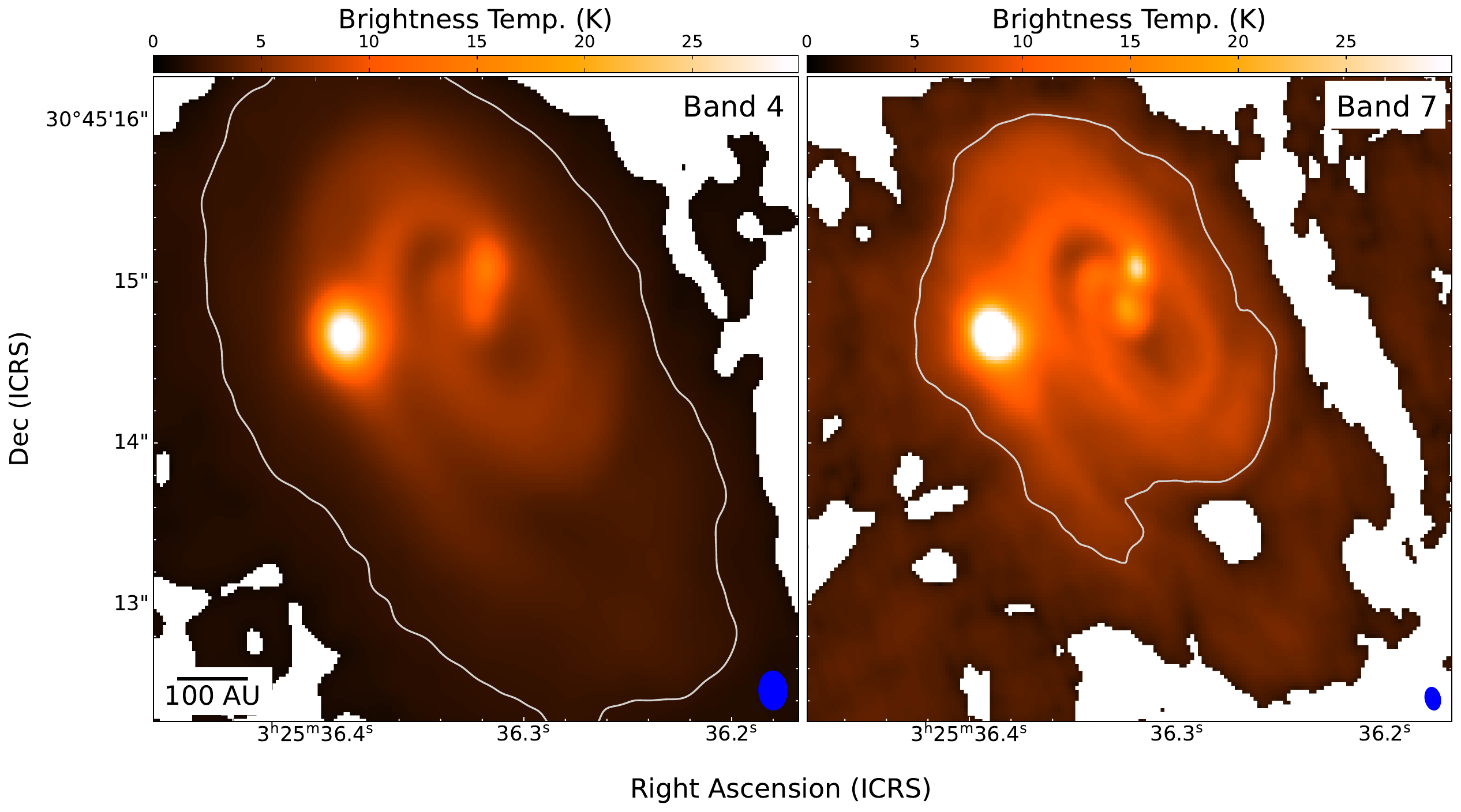}
  \caption{ALMA Band 4 and Band 7 continuum brightness temperature maps created without using the Rayleigh-Jeans approximation. Both Bands use the same colormap to ease comparison.  The light grey contour marks the 10-$\sigma$ dust continuum as an indicator of spatial scale sensitivity. The white regions outside the contour are pixels with negative values. The beams of the two maps are shown in the bottom right corners.
  }
  \label{TB}
\end{figure*}

This central ring is more noticeable in Figure \ref{TB}, which presents the Band 4 and 7 continuum brightness temperature maps (made without using the Rayleigh-Jeans approximation). The brightness temperature is the temperature required for a blackbody to match the observed intensity. In Band 7 of Figure \ref{TB}, the central ring, in which source $a$ is located, is more clearly seen. 

In general the brightness temperatures are higher at Band 7, typically by factors of 1.25 for the central region and 1.5 near the compact sources and around the east-north-west edge of the outer spiral (as traced by the 10-$\sigma$ contour in Band 7 of Figure \ref{TB}).
This is likely an optical depth effect due to the shorter wavelength Band 7 being more optically thick, which yields a brightness temperature closer to the physical temperature at its $\tau=1$ surface that should be at a higher vertical height where the temperature is likely higher.

However, the comparison between the two brightness temperatures may be affected by the observations. In particular, Band 4 is more sensitive to the envelope structure, as is evident by the substantially larger detected dust emission region in Band 4 compared to Band 7 (as suggested by the 10-$\sigma$ contours in Figure \ref{TB} and the maximum recovery scale listed in Table \ref{tab:obs}). As Band 4 has more envelope emission detected (i.e., it is less resolved out), this will decrease the relative flux of the disk in Band 7, since Band 7 observations are not on top of the envelope emission plateau.  This will also only allow lower-limits on any estimate of the spectral index, see \S \ref{sec:disc} for more discussion.

\begin{figure*}
  \centering
\includegraphics[angle=0,width=0.99\textwidth]{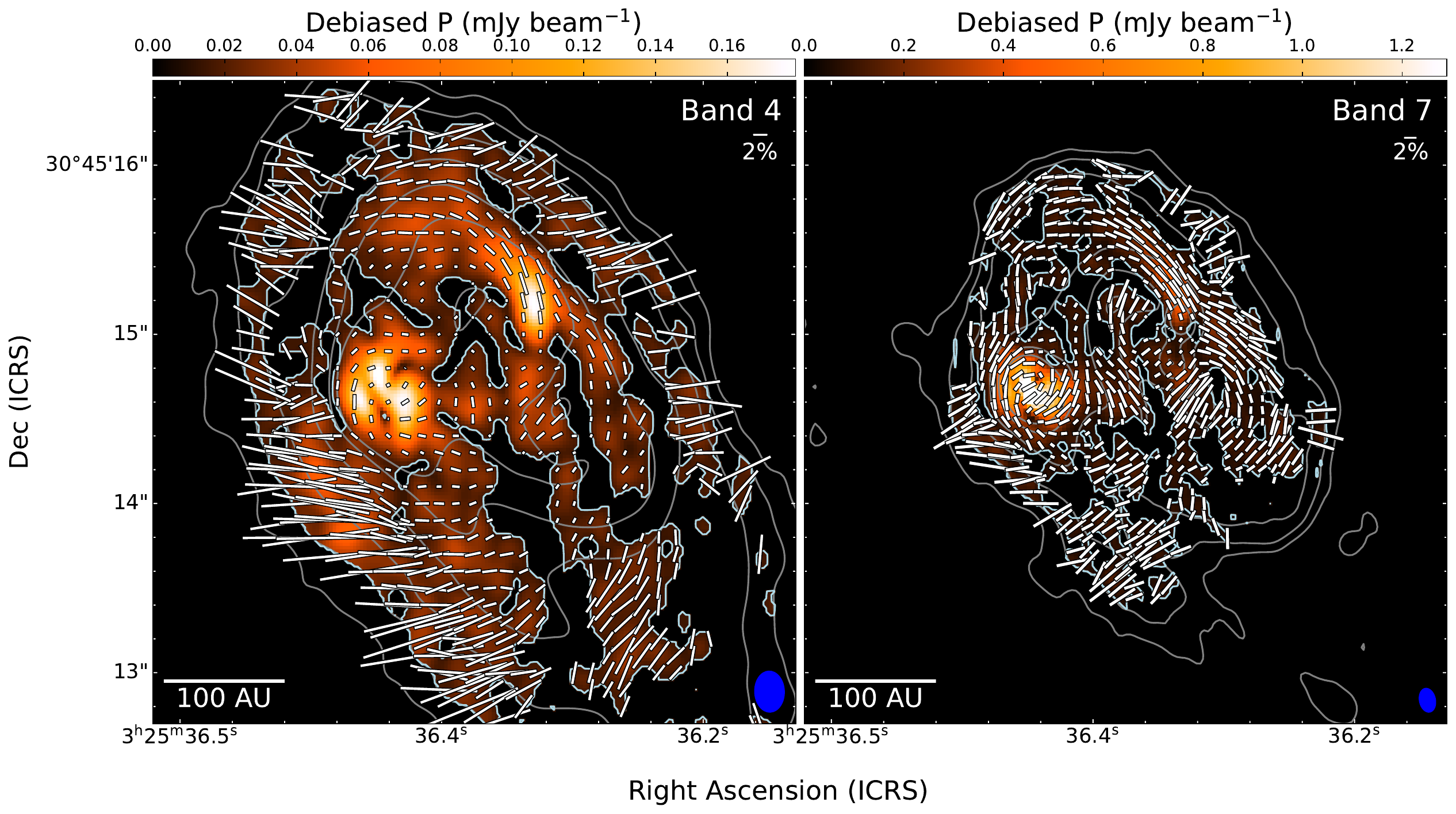}
  \caption{ALMA Band 4 and Band 7 linear polarized intensity. The light grey contours are Stokes I intensity (5, 10, 20, 50, 100, and 150 $\times\sigma_I$, where $\sigma_I$ is given in Table \ref{tab:maps}. The teal contour is the 2$\sigma_P$ level; the linear polarized intensity is only shown when P $>$ 2$\sigma_P$ and Stokes I $>$ 10$ \sigma_I$. The line segments are the polarization angle with the length related to the percent polarization; a 2\% polarization length is given in the upper right. We plot the polarization angles with Nyquist sampling along the beam minor axis. The beams of the two maps are shown in the bottom right corners.
  }
  \label{LPI}
\end{figure*}

\begin{figure*}
  \centering
\gridline{
\leftfig{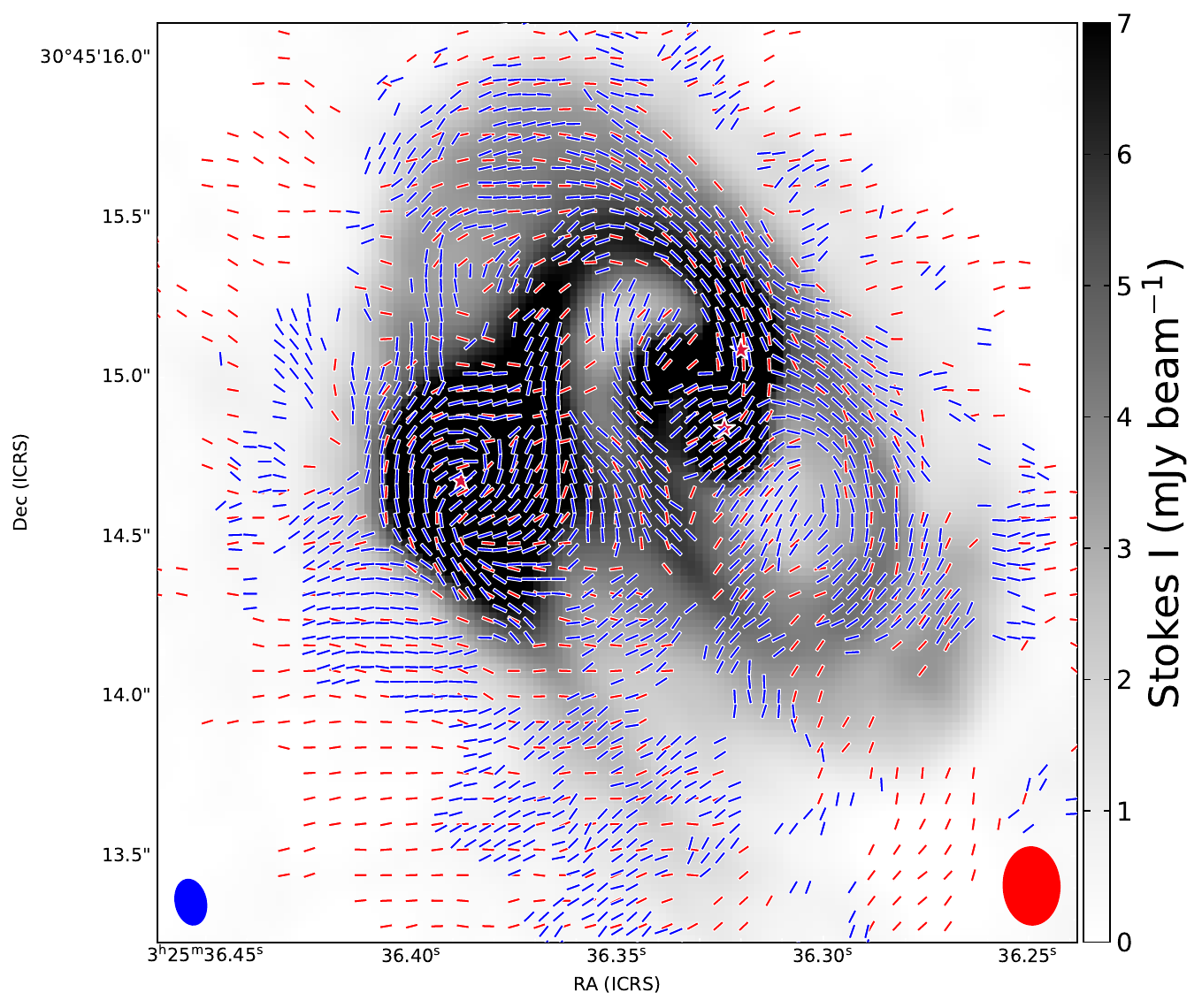}{0.5\textwidth}{}
\rightfig{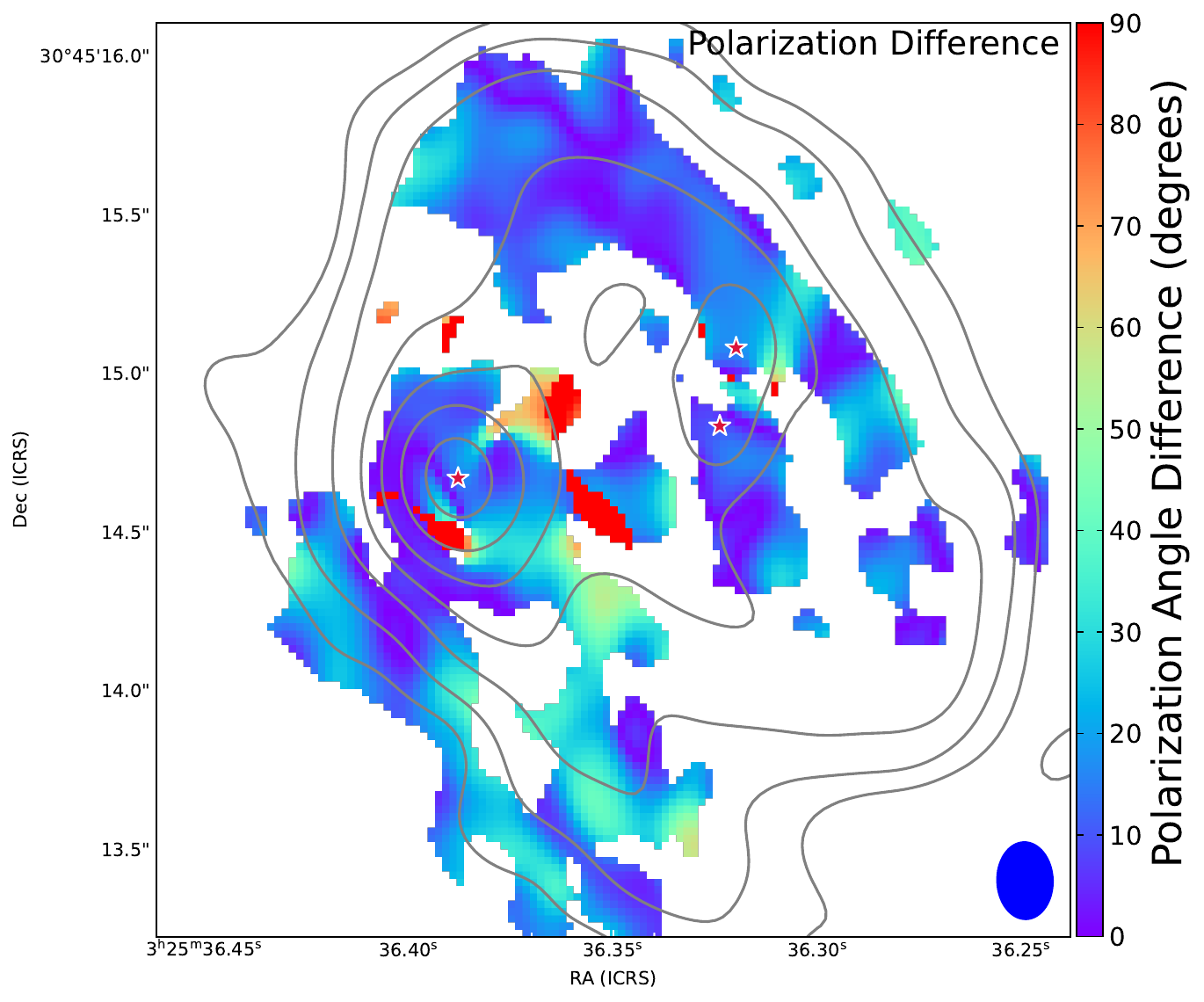}{0.5\textwidth}{}
    }

  \caption{
  Left: Band 4 (red) and Band 7 (blue) polarization angles overlaid on the Band 7 Stokes I continuum that has a linear colormap stretch to emphasize the spiral structure and the central ring in which source $a$ is located. The polarization angles are not scaled, and they are Nyquist sampled to the corresponding beam minor axis. Angles are plotted when P $>$ 2$\sigma_P$ and Stokes I $>$3$ \sigma_I$, as defined in Table \ref{tab:maps}. Crimson stars mark the location of the three compact continuum sources. The beams are shown in the bottom left (Band 7) and bottom right (Band 4).
  Right: Band 4 and Band 7 difference polarization angles overlaid on the Band 7 Stokes I continuum contours. Band 7 has been smoothed to match the resolution of Band 4. The difference angles are only plotted when both measurements have P$>$3$\sigma_P$, so the difference angles have uncertainties $\lesssim$ 13.5$^{\circ}$, and Stokes I $>$3$ \sigma_I$, as defined in Table \ref{tab:maps} for both Bands. Contours are 5, 10, 20, 40, 80, 160, and 320 $\times$ the smoothed $\sigma_I$, as defined in Table \ref{tab:maps}. Crimson stars mark the location of the three compact continuum sources. The beam for both Bands is shown in the bottom right.
  }
  \label{BothPol}
\end{figure*}

Figure \ref{LPI} presents the linearly polarized emission and polarization angle in both Bands. The polarization morphology is similar at both Bands, as is best illustrated in Figure \ref{BothPol}, which overlays the polarization angles at both Bands without any scaling for polarization fraction (Figure \ref{BothPol}, left) and the difference between the two Bands when smoothed to the same spatial resolution (Figure \ref{BothPol}, right).
It is important to note that the two wavelengths generally have the same polarization orientations in most locations in the source with a few exceptions that are likely due to differences in the \textit{u,v} sampling of the two observations or differences in optical depth. The measured polarization angles in Figure \ref{BothPol} (left) extend further out than the ones in Figure \ref{LPI} since the Stokes I cutoff is 3-$\sigma$ in the former and 10-$\sigma$ in the latter. This was done to make the polarization angles easier to view since the angles on the outskirts of the disk have very high polarization fractions (i.e. more than 10\%), which may be due to observational biases.

When observing deeply embedded protostars, the measurement of the envelope polarization is driven by the sensitivity to the larger spatial scale of the envelope structures.  With interferometeric observations, if the polarization feature of the envelope is smaller or less smooth (i.e. more clumpy) than the envelope continuum emission, one may observe envelope polarization although the envelope itself may be mostly resolved out, which then results in a very high polarization fraction since the continuum measurement is low.  This is likely the origin of the high polarization fractions seen around the outer edge of the circumbinary disk in Figure \ref{LPI} \citep[also see,][]{LeGouellec2020}.

In addition, when we compare Figures \ref{stokesI} and \ref{LPI}, the peak Stokes I locations and the peak polarization locations are in general offset.  In the case of source $a$, there is not a specific polarization peak in either Band.  For source $b$, the polarization peak is just to the north of the Stokes I peak at both Bands. For source $c$, the polarization peak is to the north of the Stokes I peak in Band 4 but to the south of the Stokes I peak in Band 7. These offsets are similar to what has been seen in other Class 0 sources such as VLA 1623 \citep{Harris2018} whereas older sources such as HL Tau or IM Lup do not have clear offsets \citep{Stephens2023,Hull2018}. This is likely due to scattering polarization from disks with large scale-heights having more polarization on the near side compared to the far-side \citep{Yang2017}, and the older disks are more settled. Indeed, recent observations of Class 0/I disks suggest that their dust is less settled than Class II disks \citep{Lin2023,Villenave2023}. However, in the case of L1448 IRS3B, it is unclear if these peak polarization offsets are due to scale-heights or some other mechanism

In any case, the polarization observed in L1448 IRS3B is  different than what is observed in other sources, either single disks or circumbinary disks. 
Morphologically, there are three notable features in the polarization seen in Figures  \ref{LPI} and \ref{BothPol}.  The most striking is that the polarized emission generally follows along the spiral structures (best seen in Figure \ref{BothPol}).  Second, around source $c$, the polarization angles wrap in an azimuthal pattern with clear nulls seen at the higher resolution of Band 7, but on the other hand, directly at source $c$ (and also the ring surrounding source $a$) there are uniformly aligned polarization angles, particularly seen in the higher resolution Band 7 observations.  Finally, around the edges of the disk, the polarization has a much higher percent polarization (i.e., $P/I$) and is aligned similarly on both sides and at both wavelengths, although more pronounced in Band 4.  
We will investigate various polarization mechanisms below to better explain the observations. 

\subsection{Other Sources in the Field}

In addition to L1448 IRS3B, there were other sources in the field.  In Appendix A we show the maps for these sources:
L1448 IRS3A, L1448 IRS3C (L1448 NW), and an unknown point source in Band 4 that is likely a background galaxy. 

\section{Discussion}
\label{sec:disc}

As discussed in \S \ref{sec:intro} above, linear dust continuum polarization observations in circumstellar disks can arise from multiple mechanisms, but in most cases in circumstellar disks to date (and mostly in older Class II sources) dust polarization is likely dominated by scattering in the shorter ALMA wavelength Bands and likely some sort of mechanical alignment in the longer ALMA wavelength Bands \citep[e.g.,][]{Gold1952,Lazarian2007b,Kataoka2019,Mori2019,Mori2021,Hoang2022,Reissl2023,Stephens2023,Tang2023,Lin2024a,Harrison2024}. However, this has not been generally true for circumbinary disks, which have been posited to have polarization due to magnetic fields in several cases \citep[e.g.,][]{Alves2018,Sadavoy2018,Harris2018,Tang2023}. Indeed, with the diversity of morphology in L1448 IRS3B, we should investigate the various polarization mechanisms more closely.

First of all, we place constraints on the optical depth of the dust emission, as this can play an important role in which type of polarization mechanism dominates \citep[e.g.,][]{Yang2017,Mori2019,Hoang2022,Lin2022,Lin2024a}.
With two wavelengths, one typically measures the spectral index ($\alpha$) to assess the optical depth of emission structures, which typically assumes optical thin emission, an incorrect assumption when observing deeply embedded protostars. However, the estimate of $\alpha$ can be even more complicated with interferometric observations of deeply embedded protostars with bright envelopes as more than beam matching is required, again for the same reasons as the brightness temperature map has systemic issues in \S \ref{IRS3B Results}, more sensitivity of the Band 4 map to larger scale structure due to different {\it u,v} sampling.  

We attempted to mitigate the difference in {\it u,v} sampling with various techniques beyond simple image smoothing to match beams, which will increase surface brightness sensitivity but not add sensitivity to structures that are not represented in the image. For example, we tried using other techniques such as (1) the {\it mtmfs}  option in tclean and (2) restricting the {\it u,v} range to where the two datasets had the most overlapping data, between 50 and 1000 k$\lambda$ coupled with a {\it u,v} taper that resulted in beams whose areas  were within a few percent of each other.  However, in both cases, the Band 4 observations were more sensitive to the envelope, which make the spiral arm structures brighter (as they are on a plateau of emission).  The bright compact sources are less affected as the envelope plateau is a smaller contamination to the total intensity. The bright compact emission had lower-limit spectral indices of $\alpha$ = 2 with the spiral arms being 1.5. A spectral index less than 2 requires high optical depth and scattering. However, due to the contamination, these are only lower limits to the spectral index. In addition, the lower brightness temperature of the spiral arms in Figure \ref{TB} also may suggest lower optical depth in the spiral arms as we are seeing deeper into the disk. In any case, to do better than these lower spectral index limits would require different observations or detailed radiative transfer modeling of a complex source, both of which are outside of the scope for this paper. 

With only weak constraints on the spectral index, we have very little knowledge of the optical depth in the source, which will make it more difficult to determine the polarization mechanisms.  Nonetheless, below we compare the different mechanisms and assess their contribution to the polarization seen in Figures \ref{LPI} and \ref{BothPol}.

\subsection{Magnetic Field Alignment}

Magnetic fields are the most common mechanism evoked for elongated dust grain alignment in astrophysics using mainly magnetic radiative alignment torques to align the dust grains perpendicularly to the field \citep[e.g.,][]{Andersson2015,Hoang2016}.  In general, the magnetic field is commonly thought to be responsible for the polarization seen in the inner envelope of embedded protostars \citep[e.g.,][]{Hull2013,Hull2014,Hull2017,Cox2018,Galametz2018,Sadavoy2018,Maury2018,LeGouellec2019,LeGouellec2020,Hull2020,Huang2024,Encalada2024} but not yet clearly in a Class 0 disk (outside of multiple systems), although there is a possible detection in a Class I disk \citep{Aso2021}. Even in the rare disk cases, other mechanism explanations and more complex models have not been exhaustively attempted. 

Previously, the highest resolution polarization observations of the inner envelope of L1448 IRS3B, which had a fractional polarization of 2.7\% \citep{Galametz2018}, were argued to be due to magnetic field alignment. In that case, the field has a direction that is approximately perpendicular to the east-west outflow \citep[see Figure 2 of][]{Galametz2018} that mostly originates from source $c$ \citep{Nick2021}. 

As already mentioned, our observations at higher resolution show a much more complicated morphology than the previous polarization observations.  Nonetheless, the outer disk region polarization in Figure \ref{LPI} (i.e. around the edges in the Band 4 maps) is in general consistent with the inner envelope polarization angles from \cite{Galametz2018} (-67$^\circ$).  The percent polarization for the inner envelope polarization is higher than the low resolution observations, at 8\% in Band 4 and Band 7, even higher at the edges of Band 4. The reason for the high polarization fraction is, as discussed above, likely due to the partially resolved out Stokes I envelope, especially in Band 4 (compare the two images of Figure \ref{LPI}), although an intrinsically higher polarization fraction in the inner envelope than in the disk is also possible (indeed expected to some extent),  as indicated by, e.g.,  the ALMA dust polarization survey of Perseus and Orion protostellar systems \citep{Cox2018,Huang2024}.

\begin{figure*}
  \centering
    \includegraphics[angle=0,width=0.70\textwidth]{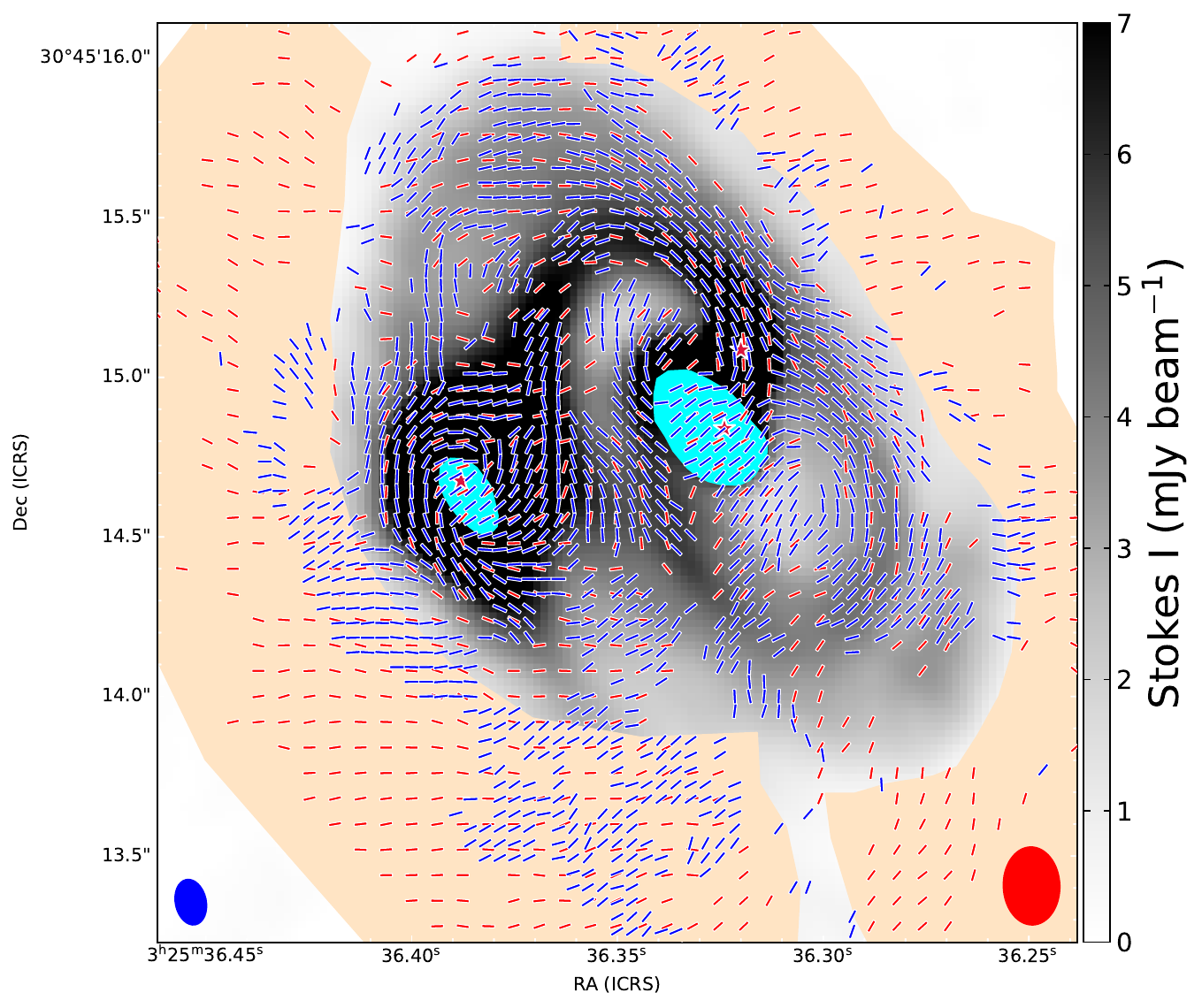}
  \caption{
  Same as Figure \ref{BothPol} (left) but with two regions highlighted for likely magnetic fields (tan) and likely scattering (cyan).
  }
  \label{Regions}
\end{figure*}

In Figure \ref{Regions}, we highlight different regions of polarization.  Those highlighted in a tan color are those we posit are generally consistent with the low-resolution observations, coming from the inner envelope and likely due to magnetic field dust grain alignment.  The region was chosen using a combination of polarization fraction and location (off the circum-multiple disk). There are regions to the north and south that are ambiguous in both Bands so they were not included in the posited region.  In addition, there are regions in the circum-multiple disk, particularly to the southwest of source $c$ (i.e., bottom middle of Figure \ref{Regions}), that are generally comparable to the low resolution angles so may also be associated with the magnetically aligned dust grains.  

The significant difference between the lower resolution observed morphology and the higher resolution observation morphology is that the outer region (around the circum-multiple disk) polarization generally trends radially toward the the system center. This is best seen in Band 4 in Figures \ref{LPI} and \ref{BothPol} (left) in the upper left of the Figures. It is important to note that magnetic field alignment results in polarization that is rotated by 90$^\circ$ to the magnetic field of the plane of the sky, which in this case would suggest a somewhat toroidal field morphology (i.e. the polarization direction is for the most part radially oriented). 
To show this possible field orientation, we also display Figure \ref{Regions} with the polarization angle rotated by 90$^\circ$ for the possible inferred magnetic field in the plane of the sky in the Appendix (Figure \ref{RegionsRotated}).

A combination of toroidal and poloidal fields are commonly evoked in simulations of disks \citep[e.g.,][]{Balbus1991} while observationally toroidal fields are typically consistent with the observations at the inner envelope scale \citep[e.g.,][]{Kwon2019}. We posit that these highlighted observations also may be  generally consistent with a predominantly toroidal field in the inner envelope, although there are many vectors that are not in agreement with a pure toroidal field.

Inside of the regions with high polarization fraction in Figure \ref{LPI}, the polarization morphology in many cases switches dramatically, frequently with $\sim$90$^\circ$ rotation, and the mean polarization fraction drops to 1.2$\pm$0.6\% and 2.4$\pm$1.1\% in Band 4 and Band 7, respectively.  In addition, there is an azimuthal morphology around source $c$ and a general alignment of the polarization with the spiral arms, specifically seen in Figure \ref{BothPol} (left). In both of those cases, it would be difficult to suggest that the polarization is due to magnetic field alignment as the implied magnetic field direction would be radial around source $c$ and perpendicular to the spiral arm structures, both of which are not expected field morphologies since the magnetic field is expected to be wrapped by differential rotation into a predominantly toroidal configuration around source $c$ and an orientation approximately parallel to the spirals.

If the dust grain sizes are comparable to the observing wavelength, it has been shown that the polarization could be along the minor axis of the dust grains, perpendicular to the often adopted orientation \citep{Kirchschlager2019}. For the multi-wavelength polarization observations toward BHB[2007]-11 \citep{Alves2018}, \cite{Guillet2020} used this polarization flip as a possible explanation for the observations using a dust model with 1 mm dust grains in a log normal distribution and suggested that the actual magnetic field may be dominated by a toroidal field instead of poloidal, assuming the grains are aligned by magnetic fields. 
The inner region of the L1448 IRS3B circum-multiple disk has similar spiral structures with polarization aligned along them; if large dust grains exist in the disk and there are magnetic fields along the spirals, then it is possible to match the observed polarization morphology using a similar model.
However, the polarization in the Guillet et al. (2020) model also predicts that the Band 4 polarization fraction is larger than Band 7 (see their Figure 1) which, while consistent with the data for BHB[2007]-11, is inconsistent for L1448 IRS3B. The Guillet et al. (2020) model used a very specific dust size distribution, grain aspect ratio, and angle between the grains’ minor axis and the line-of-sight for the BHB[2007]-11 fit, so more modeling is needed to draw firm conclusions.
However, as the existence of large dust grains (i.e., $\sim$1~mm) in the circum-multiple disk of L1448 IRS3B is unclear and as the Band 4 polarization fraction is smaller than Band 7, we posit that the magnetic alignment of dust grains in the spiral structures is less favorable than mechanical alignment (see discussion below). 


\subsection{ Polarization from Self-Scattering}

When dust grains are comparable to the observing wavelength, the dust grains can self-scatter their thermal emission creating a linear polarization continuum on the order of 1\% \citep[e.g.,][]{Kataoka2015}. The scattering polarization mechanism results in a distinctive polarization angle morphology in an inclined disk with the polarization direction being parallel to the disk minor axis
\citep{Kataoka2016,Yang2016}.  This morphology is typically seen in disks in the shorter wavelength ALMA Bands, especially Band 7 \citep[e.g.,][]{Stephens2014,Segura-Cox2015,Hull2018,Bacciotti2018,Dent2019,Harrison2024}. 

However, when disks are observed at longer wavelengths (e.g., ALMA Band 3), the polarization mechanism is often, but not always, dominated by another mechanism that changes the observed polarization morphology, likely due to a decrease in optical depth, which reduces the contribution from self-scattering and allows the polarized emission from other mechanisms to dominate without being excessively extincted \citep[e.g.,][]{Kataoka2017,Harris2018,Mori2019,Mori2021,Lin2022}. 

Although, this shift in polarization morphology is commonly observed in disks, there are also sources where the polarization morphology is consistent with scattering at all ALMA Bands \citep[e.g., MWC 480 and RY Tau,][]{Harrison2019}. On the other hand, polarization due to dust self-scattering is highly dependent on the maximum dust grain size, falling quickly with increasing wavelength \citep[e.g.,][]{Kataoka2015}.  In fact, to match the polarization observations of MWC 480 and RY Tau, \cite{Harrison2024} require either high optical depth or, in the case of MWC 480, segregation of a large grain distribution (near the midplane) and small grain distribution (above and below the midplane).

In the case of L1448 IRS3B, the overall circum-multiple disk does not generally exhibit a polarization direction that matches a scattering morphology since the majority of the polarization vectors are not aligned with the minor axis of the circum-multiple disk nor is the polarization direction uniform. However, there are two regions in the circum-multiple disk that broadly match the polarization morphology aligned with a minor axis for scattering dominated polarization: around source $c$ and the disk/ring of the central source, called source $d$ in \cite{Reynolds2023}, see Figure \ref{Regions}, cyan.  These two regions do have polarization that is generally aligned with the circum-multiple disk minor axis. 

In \cite{Reynolds2023} at higher resolution, they fit source $c$ with a Gaussian in the \textit{u,v} plane and find an extended source of 0$\farcs$2 $\times$ 0$\farcs$175 with a PA of 25$^\circ$. The average dust polarization direction in Band 7 of the region around source $c$, highlighted in Figure \ref{Regions}, has an angle of 126$^\circ$, which is comparable to being along the minor axis of the compact source $c$ emission.  This type of polarization morphology shift to scattering-like dominant on the smaller scale is consistent with other sources, such as VLA 1623 \citep{Harris2018,Sadavoy2018} and many Perseus protostars \citep{Cox2018}.

Similarly, in \cite{Reynolds2023} they note a dust ring around source $d$ with an estimated PA of 45$^\circ$ PA. The average dust polarization direction in Band 7 of the region around source $a$, highlighted in Figure \ref{Regions}, has an angle of 220$^\circ$, which is comparable to being along the minor axis of the ring emission.  Note that the region highlighted is centered on source $a$ and not source $d$. The polarization emission around source $c$ is much higher than the polarization emission around source $a$ (see Figure \ref{LPI}), which may suggest differences in dust grain sizes that might be expected if the source $c$ structure formed recently from more evolved dust through fragmentation of a gravitational unstable circum-multiple disk \citep{Reynolds2023}.

\subsection{Radiative Alignment}

When dust grains spin with a precession rate from radiative torques that is faster than the Larmor precession rate around the magnetic field then the dust grains can align with an anisotropic radiation field  \citep{Lazarian2007,Tazaki2017}.  For dust in disks, this means that the dust grains have their long axis aligned perpendicular to the local radiation flux, i.e. the direction of the local radiation anisotropy, which results in an azimuthal polarization pattern.  At first glance, this mechanism was thought to explain the polarization morphology detected in circumstellar disks at the longer wavelength ALMA Bands, but this is likely incorrect.  Radiation alignment polarization produces a circular polarization morphology at all disk inclinations and not the elliptical morphology that is detected \citep{Yang2019}.  

The morphology of radiative dust alignment is not a good match for the L1448 IRS3B polarization. Although the radiation field is much more complicated than a single system, the large scale polarization curvature that one sees along the spiral arm structures is not circular from any collection of radiation sources.  On smaller scales, the most noticeable circular or elliptical polarization morphology is the polarization surrounding source $c$, away from the central region. 
However, that polarization morphology is not circular (as expected for radiative dust alignment) either; it is very elongated or elliptical, which is inconsistent with radiative dust alignment. In other words, we do not see strong evidence of radiative dust alignment in L1448 IRS3B.

\subsection{Mechanical Alignment}

As discussed above, azimuthally aligned effectively prolate grains are needed to explain the longer ALMA Band polarization morphology observations around most circumstellar disks \citep{Kataoka2017,Yang2019,Mori2019,Harrison2019,Ohashi2023,Lin2022}. In HL Tau, for example, most wavelength observations require a mixture of azimuthally aligned, prolate dust grains with self-scattering \citep{Lin2024a}. The mechanism that is aligning the dust grains is still not known, but it is thought to be similar to the Gold mechanism \citep{Gold1952}, where dust grains are aligned mechanically by the relative difference in the gas and dust motion, but without the requirement for supersonic relative velocities. The current best models for this are the v-MET and Badminton Birdie alignment mechanisms \cite[e.g.][]{Lazarian2007,Hoang2016,Hoang2018,Lin2024b}. In any case, without knowing the details of the physical process, studies have suggested that such alignment, along with dust scattering, can explain the multi-wavelength polarization observations toward HL Tau \citep{Lin2024a}, as well as the high-resolution HL Tau observations with resolved rings and gaps at Band 7 \citep{Stephens2023}. In the latter example, the models specifically require effectively prolate dust grains.

In the case of L1448 IRS3B, the dust is not generally aligned azimuthally as found in single sources like HL~Tau, likely due to multiple sources and the complicated spiral arm structure of the circumbinary disk. Nonetheless, polarization in both Bands demonstrate smaller-scale azimuthal morphology around source $c$ and, more remarkably, show an overall alignment with the spiral arm structures.

Around the vicinity of source $c$, the polarization morphology at Bands 4 and 7 broadly resembles the azimuthal morphological transition of HL Tau. As mentioned earlier, near the source center (cyan region in Figure \ref{Regions}), the polarization at both Band 4 and 7 appears parallel to the disk minor axis with the shorter wavelength exhibiting a larger polarization fraction, which is consistent with scattering. However, near the outskirts of source $c$ (within 2~beams outside of the cyan region in Figure \ref{Regions}), the polarization morphology changes and becomes azimuthally oriented for Band 7. For Band 4, only the southeast portion appears consistent with azimuthally oriented polarization. Nonetheless, both bands show two polarization holes that are almost along the disk minor axis, see the Figure \ref{BothPol} colormap around source $c$, at slightly different locations between Band 4 and 7. The change of these features from the shorter to the longer wavelength, and from near the protostar to farther away, may be explained by decreasing the optical depth of scattering of azimuthally aligned, prolate grains \citep{Lin2022, Lin2024a} and a difference in the beam resolution. A drop in polarization from Band 7 to Band 3 or 4 at the center of the disk was also observed in HL Tau \citep{Stephens2017, Lin2024a}. 

A notable difference between the polarization from source $c$ and HL Tau is in the orientation of the polarization holes. For perfectly azimuthally aligned prolate grains, the polarization holes should be along the circumstellar disk minor axis where the scattering polarization cancels out the thermal polarization from prolate grains whose long axes are parallel to the circumstellar disk major axis. 

However, the polarization holes around source $c$ in Band 7 (Figure \ref{LPI}) appear rotated by $\sim 20^{\circ}$ east-of-north. Such a rotation can be explained by the scattering of prolate grains aligned azimuthally but with an inward spiral toward the center (\citealt{Lin2022}; see their Figure 16), surrounding the source $c$ disk. Based on this rotated direction, we can infer that the alignment direction of the prolate grains should be clockwise and inward towards source $c$. The direction coincides with the direction of rotation of the main disk \citep[clockwise rotation and source $c$ is positioned in the near-side of the main disk;][]{Nick2021}. The observed polarization morphology inspiral behavior around the disk of source $c$ is consistent with polarization morphological spirals from two other sources, AS 209 and GG Tau \citep{Mori2019,Tang2023}. Following the arguments from those two sources, the inspiral polarization could trace the inward accretion flow onto the central source if the alignment mechanism is the same for all three targets.

As discussed above, the polarization near the vicinity of source $c$ has an azimuthal morphology that is similar to the observations of an isolated source like HL Tau. From that similarity, we suspect that the underlying alignment mechanism must be dominated by some physical property or condition near source $c$, which includes the possibility of a Keplerian flow with accretion, as is suggested in HL Tau \citep{Lin2024a}. 
However, the northwest portion of the disk from source $c$, in both Bands but particularly clear in Band 4, does not resemble the curvature expected from azimuthally aligned prolate grains. Given that the polarization angles curve toward sources $a$ and $b$, the alignment direction could be ``disturbed'' by influences beyond source $c$. This may be consistent with the fact that source $c$ is not as massive compared to sources $a$ and $b$ \citep[$\leq$0.2~M$_\odot$ for source $c$ and 1.19~M$_\odot$ total for sources a and b;][]{Nick2021}.

Beyond the vicinity of source $c$, the most remarkable aspect of the observations is that the polarization patterns closely align with the spiral arm structures. The polarization directions at both wavelengths are similar and are generally aligned with the Stokes~I arc (Figure \ref{BothPol}). 
This is suggestive that the underlying alignment mechanism may be dominated by the motion of the flows and disk material. Using the Stokes~I continuum, we can create a simple model to compare the spiral arms with the polarization to make a more direct comparison.

\subsubsection{Simple Model of Polarization Morphology}

Under the assumption that the spiral arms are creating a drift velocity of the dust relative to the gas (and traced by the dust continuum), we can make a simple model of the polarization morphology just using the spine of the continuum spiral structure. The polarization may be due to v-MET \citep{Lazarian2007}, Badminton Birdie alignment \citep{Lin2024b}, or another, yet unknown, alignment mechanism.
To compare the polarization direction with the spiral structures, we need to define the spiral structure more quantitatively. 

\begin{figure}
  \centering
\includegraphics[angle=0,width=0.50\textwidth]{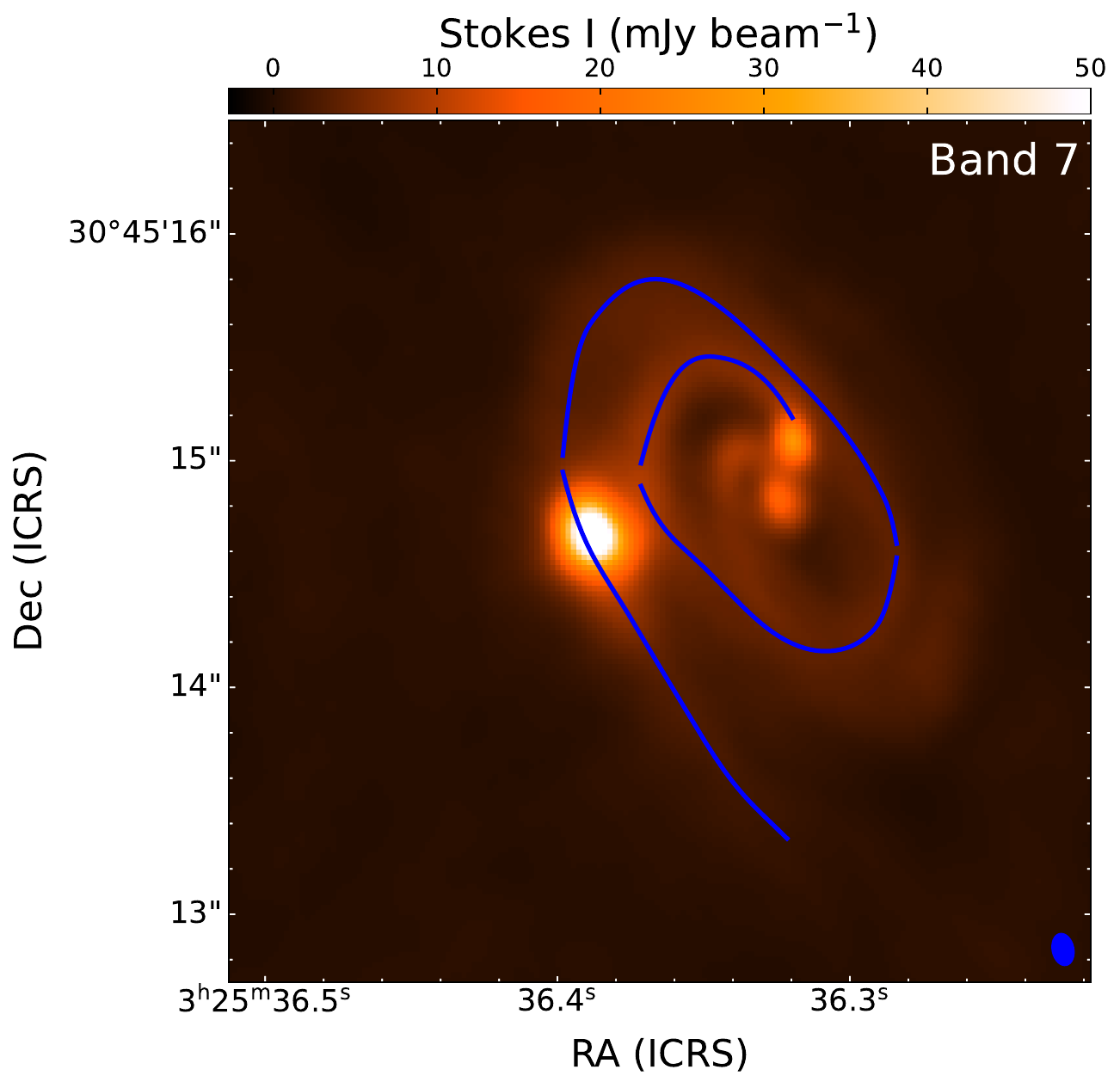}
  \caption{
  Same as Figure \ref{stokesI}, but with a spline fit of the continuum spiral structures.  The blue lines are the 4 spline fit segments.
  }
  \label{splinefits}
\end{figure}

We use a simple mask to define the spiral structures in the Stokes I maps, then fit cubic splines to the emission using \texttt{UnivariateSpline} in \texttt{SciPy} with a smoothing factor = 5. Using Band 7, we show the results of this fit in 4 segments in Figure \ref{splinefits}. The 4 segments are required due to the \texttt{UnivariateSpline} constraint of monotonically increasing input values. Band 4 gives a similar result, but we use Band 7 since the higher resolution provides a more accurate fit to the substructure (see Figure \ref{stokesI}).

\begin{figure}
  \centering
\includegraphics[angle=0,width=0.50\textwidth]{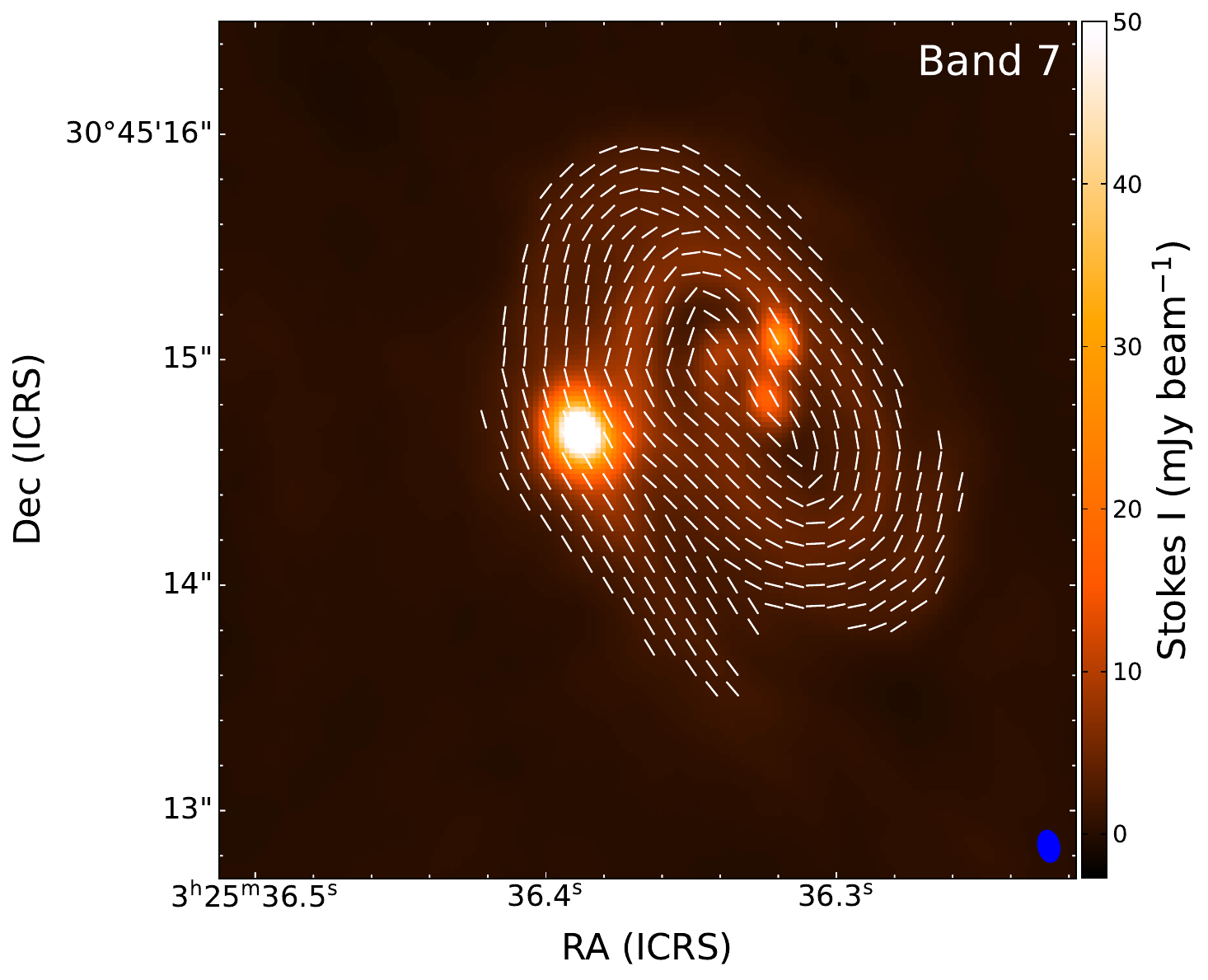}
  \caption{
  Model of the polarization angles derived only using the spiral arm structures fit in Figure \ref{splinefits} and as described in the text.
  }
  \label{modelvectors}
\end{figure}

For every pixel in the Stokes I map, we use the nearest fitted spline pixel to calculate that pixel's polarization vector, which is essentially the arctangent of the instantaneous slope of the spline fit. The slopes are obtained via the \texttt{UnivariateSpline} computed derivative attribute. The results for Band 7 are shown in Figure \ref{modelvectors}. The vectors of the model are very similar to the observations without shading in Figure \ref{Regions}.

\begin{figure}
  \centering
\includegraphics[angle=0,width=0.50\textwidth]{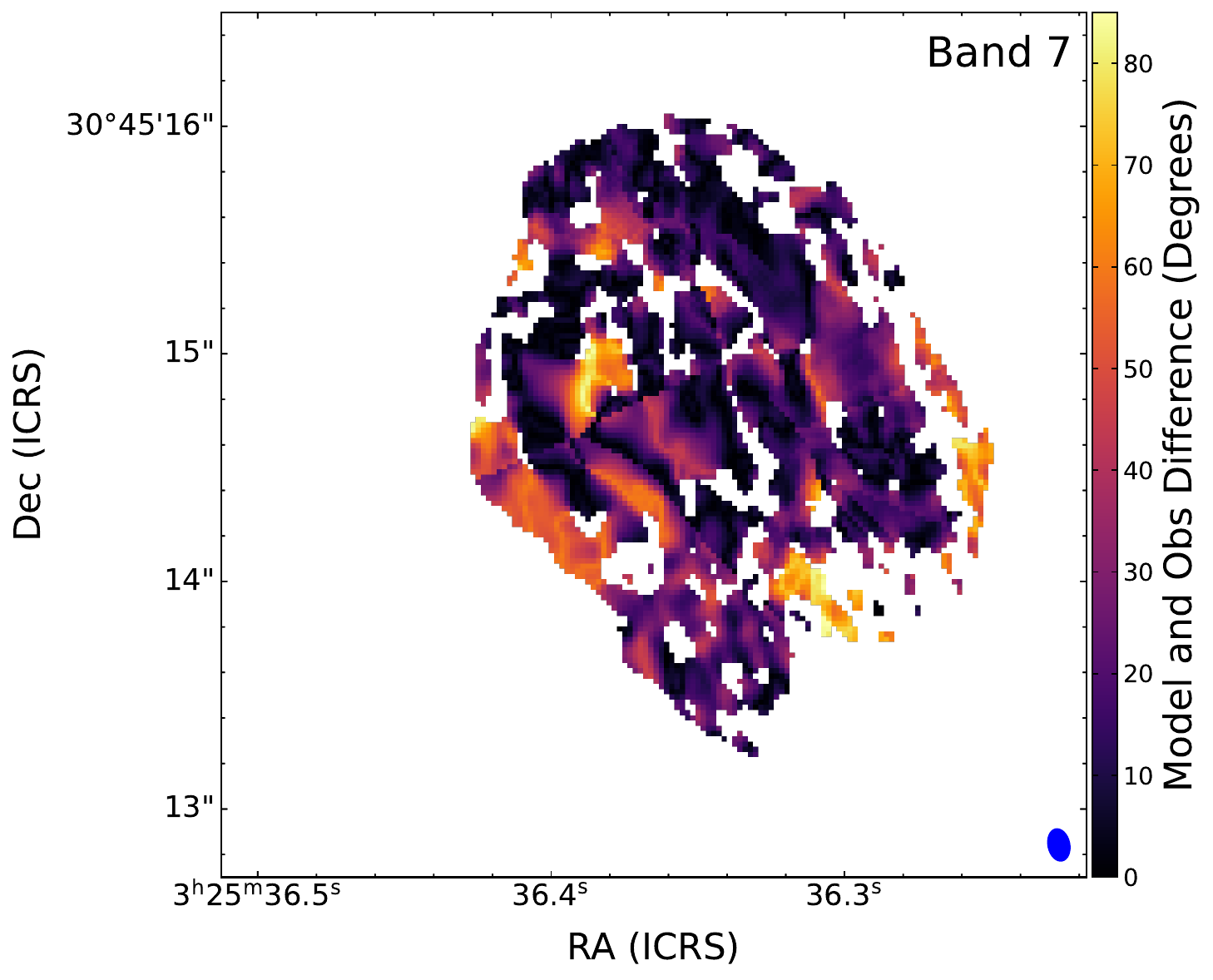}
  \caption{
  Map of the difference between the angle predicted by the simple model and the observations at Band 7 for overlapping pixels where P $>$ 2$\sigma_P$ and Stokes I $>$ 10$ \sigma_I$.
  }
  \label{ModelDiffMap}
\end{figure}

\begin{figure}
  \centering
\includegraphics[angle=0,width=0.50\textwidth]{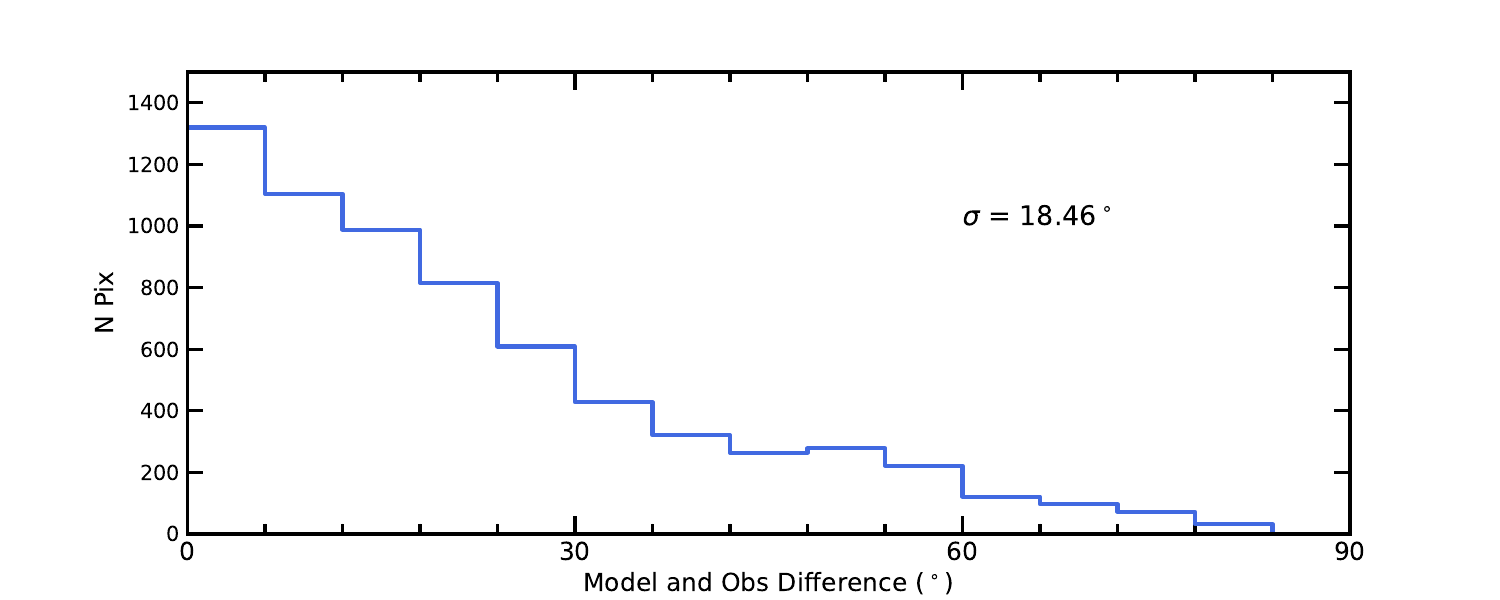}
  \caption{
  The histogram of the absolute differences between the model and the observations in Band 7 for overlapping pixels where P $>$ 2$\sigma_P$ and Stokes I $>$ 10$ \sigma_I$.
  }
  \label{ModelHisto}
\end{figure}

We can further quantify the comparison by taking the difference between the model in Figure \ref{modelvectors} with the observations for Band 7.  Figure \ref{ModelDiffMap} shows the difference angle for all pixels and Figure \ref{ModelHisto} the histogram of those differences. In Figure \ref{ModelHisto}, the standard deviation of the distribution is 18.5$^\circ$ with the distribution centered around 0$^\circ$, both of which are good for such a simple model. Thus, in both Bands, this simple model can broadly describe the observations, which implies that the polarization mechanism could be a gas-flow alignment of the dust grains.

We measure the mean fractional polarization along the spiral arms (one measurement per beam along the feature shown in Figure \ref{splinefits}), 
 1.5$\pm$0.7\% and 2.5$\pm$1.3\%, for Band 4 and Band 7, respectively, which are consistent with the values in the inner region overall. This again suggests that using the model of \cite{Guillet2020}, discussed above, would not be appropriate for this source.

\subsubsection{Model Implications}

This polarization alignment with the dust continuum spiral structure in L1448 IRS3B has not been shown before. If the dust grain alignment mechanism traces the gas flow, the polarization suggests a possible flow along the spiral arm. 
The inward spiral angle of the polarization is in the same direction as the spiral angle from polarization in the vicinity of source $c$ which hints at a common alignment mechanism, likely of mechanical origin, at least in these two regions. While the kinematic motion traced in C$^{17}$O (at only a slightly lower angular resolution of 0$\farcs$2) is largely Keplerian without any clear deviations \citep{Nick2021}, the C$^{17}$O emission is possibly not tracing the dense spiral structures. 
Further investigation into the system kinematics is warranted. 
Simulations of spiral structure in a gravitational unstable disk, such as L1448 IRS3B \citep{Nick2021}, show that once the dust drifts into the spiral arms, the dust will move along the spiral arms \citep{Rowther2024}, which may explain the observations. 
In other words, if there are spiral arms in the gas, the dust should drift to the pressure maxima, then once trapped at the pressure bump, the dust will move along the spiral arms.
If the dust polarization is tracing the dust drift, such observations may provide a direct probe of the kinematics on small scales that may be difficult to trace with molecular tracers.

The polarization fraction in the main disk is between 1\% to $2\%$, which is similar to HL Tau at ALMA wavelengths and at low resolution \citep{Stephens2017, Lin2024a}. However, high-resolution Band 7 and long wavelength (VLA Q-Band at $\lambda=7.1$~mm) observations of HL Tau suggest an intrinsic polarization of $\sim$10\% \citep{Lin2024a}. The lower level of polarization at the lower resolution in the mechanically aligned dust grains is thought to be due to beam averaging of optically thick substructure that lowers the inferred intrinsic polarization. 

With the current observation of L1448 IRS3B, it is unclear if the percent-level polarization is also due to a similar resolution effect or an actual lower level of intrinsic polarization. The latter scenario is particularly interesting since the decreased polarization can be due to less elongated grains or poorer alignment efficiency, which may be reasonable given the highly dynamic conditions in a multiple stellar system compared to an isolated system.

As mentioned previously, there appears to be a sharp transition from the outer region with high polarization to the disk region with lower polarization and a dramatic switch in polarization angle (Figure \ref{Regions}). The transition appears to coincide where the Stokes~$I$ increases as part of the main disk. We suspect that the switch in polarization is because the density rapidly increases, which prevents alignment with the magnetic field but benefits mechanical alignment.  Alignment with the magnetic field (determined by the Larmor precession timescale) needs to occur faster than the gaseous damping timescale \citep[e.g.][]{Tazaki2017} and the high-density conditions in disks easily shorten the gas damping time \citep{Lam2021}. This implies that mechanical alignment, which relies on gas bombardment to provide alignment, should be more efficient in higher-density regions.




\section{Conclusions}

We present ALMA dust continuum linear polarization observations of 
the proto-multiple source L1448 IRS3B in Band 4 and Band 7, $\lambda$ = 2.1 mm and 873 $\mu$m, respectively. We detect the polarization in the inner envelope, circum-multiple disk, spiral structures, and compact structures around individual stellar components. The circum-multiple circumstellar disk is clearly complex and provides important clues into the formation of young, close multiple protostellar systems. With polarization observations, we open a new window into the physics of multiple protostellar formation. In particular, it is surprising to find a source with multiple polarization mechanisms (as suggested by the polarization morphology), so exploring this source, and others like it, provide a unique view of star formation. The main conclusions from these observations are:

\begin{itemize}

\item The dust continuum is consistent with previous observations of the circum-multiple disk, the spiral structures and the compact sources \citep{John2016,Nick2021,Reynolds2023}.  This is the first time that the spiral structures have been resolved in Band 4.

\item The brightness temperatures of the dust in the two wavelength bands are very consistent with slightly higher temperatures for Band 7, indicating that Band 7 is less optically thin than Band 4.  The estimation of optical depth is hampered by differences in the \textit{u,v} sampling. 

\item The linear dust polarization morphology is consistent with each other across both Bands. In single circumstellar disk systems, the polarization often changes morphology across the ALMA Bands, but in this source the morphology does not present much deviation in polarization angle.  Although this is seen in a few single sources \citep[e.g., MWC 480 and RY Tau,][]{Harrison2024} and binary sources \citep[e.g., BHB07-11,][]{Alves2018}, it is suggestive of a single type of dominating polarization in each region with a weak optical depth dependence for most of the polarized emission.

\item On the other hand, the linear dust polarization morphology is different in the different structures of L1448 IRS3B.  In the inner protostellar envelope, the observations are possibly due to dust grains aligned by magnetic fields, which could be due to a generally toroidal field surrounding the circum-multiple disk.  Around source $a$ and $c$, the observations are possibly due to polarization from dust self-scattering, which suggest high optical depth.  Along the spiral arms, the observations are possibly due to polarization arising from mechanical alignment of the dust grains due to the relative difference in the velocities of the gas and dust.  In the last case, this would suggest that the local dust drift velocity field is along the spiral structures. 

\item L1448 IRS3B is one of the few sources with more than 2 polarization mechanisms occurring at the same time, making it an ideal testbed for modeling, although the complication of the triple or quadruple system makes this very challenging.  In addition, the possible detection of mechanical alignment with the gas flows along the spiral structures offers a unique probe of the relative gas and dust flows in these young multiple systems that needs to be better understood.

\item Finally, polarization is also detected toward L1448 IRS3A and L1448 IRS3C (L1448 NW), and a point source is detected in Band 4 that is likely a background galaxy. 

\end{itemize}


L.W.L. acknowledges support from NSF AST-1910364 and NSF AST-2307844.
Z.-Y.D.L. acknowledges support from NASA 80NSSC18K1095, the Jefferson Scholars Foundation, the NRAO ALMA Student Observing Support (SOS) SOSPA8-003, the Achievements Rewards for College Scientists (ARCS) Foundation Washington Chapter, the Virginia Space Grant Consortium (VSGC), and UVA research computing (RIVANNA).
Z.-Y.L. is supported in part by NASA 80NSSC20K0533, NSF AST-2307199, and the Virginia Institute of Theoretical Astronomy.
W.K. was supported by the National Research Foundation of Korea (NRF) grant funded by the Korea government (MSIT) (RS-2024-00342488). 
We thank Eli Sofovich for early data analysis.

This paper makes use of the following ALMA data: ADS/JAO.ALMA\#2021.1.00276.S. ALMA is a partnership of ESO (representing its member states), NSF (USA) and NINS (Japan), together with NRC (Canada), MOST and ASIAA (Taiwan), and KASI (Republic of Korea), in cooperation with the Republic of Chile. The Joint ALMA Observatory is operated by ESO, AUI/NRAO and NAOJ. The National Radio Astronomy Observatory is a facility of the National Science Foundation operated under cooperative agreement by Associated Universities, Inc.

The final maps and the scripts used to make them (as well as the selfcal script) are available in the Illinois Data Bank \citep{illinoisdatabankIDB-5442905}. The ALMA data and reduction scripts are available from the ALMA archive. 


%

\vspace{5mm}
\facilities{ALMA}


\software{astropy \citep{2013A&A...558A..33A,2018AJ....156..123A,Astropy2022} and SciPy \citep{SciPy}.
          }



\appendix

\section{Other sources in the Field}

\begin{figure*}
  \centering
    \includegraphics[angle=0,width=0.90\textwidth]{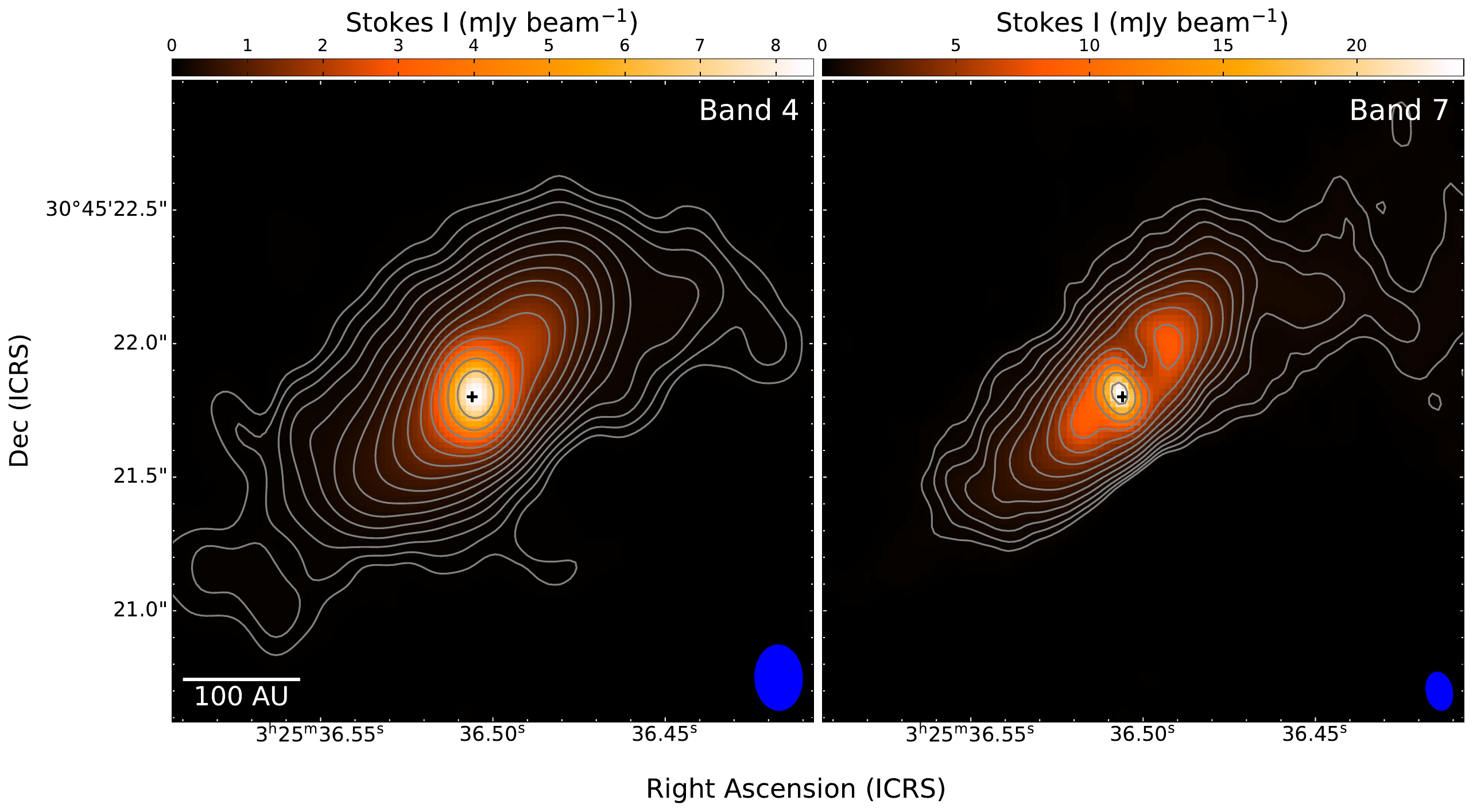}
  \caption{Band 4 and Band 7 continuum maps of L1448 IRS3A. 
  The contours are 3$\sigma\times(2)^{n/2}$, with integer n starting at 0 and using the $\sigma$ listed in Table \ref{tab:maps}.
  The beams, sizes listed in Table \ref{tab:maps}, are shown in blue in the bottom right corners. The black + symbol is the source location from \cite{Reynolds2023}.
 }
  \label{stokesI_IRS3A}
\end{figure*}

Figure \ref{stokesI_IRS3A} presents the L1448 IRS3A dust continuum (Stokes I) at the two wavelengths. L1448 IRS3A is a likely Class I source that is $\sim$7\arcsec\ to the northeast of L1448 IRS3B \citep{Looney2000,Pokhrel2023,Reynolds2023}. It is a wide-binary with a clear dust disk well detected that is consistent with the disk detected in \cite{Nick2021} and further resolved into a ring and a compact source in \cite{Reynolds2023} that suggests a disk with a large gap.

\begin{figure*}
  \centering
    \includegraphics[angle=0,width=0.50\textwidth]{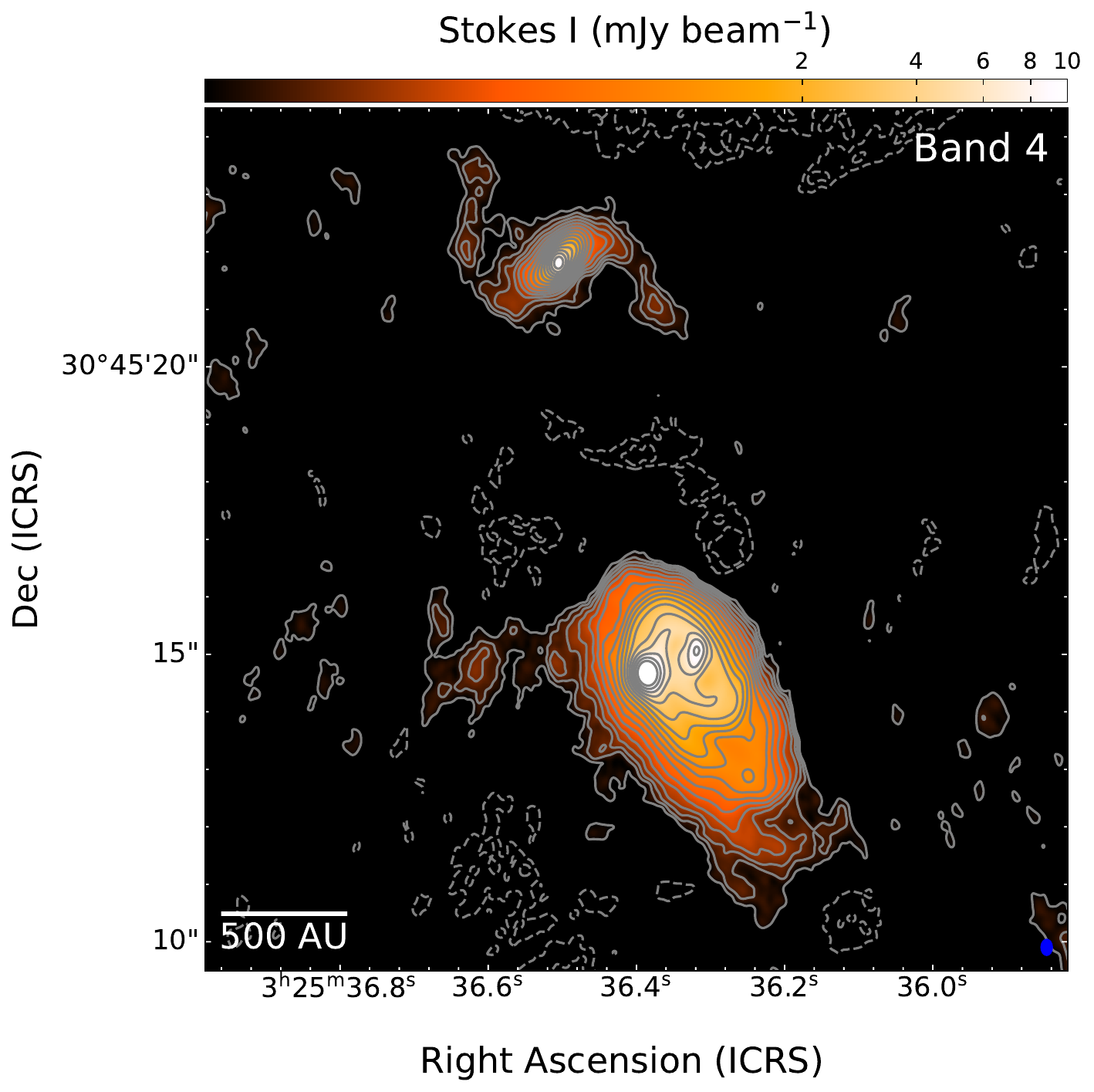}
  \caption{ALMA Band 4 continuum map of L1448 IRS3A and IRS3B using robust = 1.0. This weighting was used to better emphasize the larger scale emission, particularly to show the low level spiral arms of L1448 IRS3A.  
  The contours are $\pm$2 and $\pm$3$\sigma\times(2)^{n/2}$, with integer n starting at 0 and using the $\sigma$ listed in Table \ref{tab:maps}.
  The beam, size listed in Table \ref{tab:maps}, is shown in blue in the bottom right corner.
 }
  \label{stokesI_IRS3-lowres}
\end{figure*}

Furthermore, Figure \ref{stokesI_IRS3-lowres} is a zoom out of the emission, showing both L1448 IRS3A and IRS3B in Band 4 only (since it has a larger spatial scale sensitivity). We used Robust weighting of 1.0 as a compromise between sensitivity and larger spatial scale sensitivity.  IRS3A has clear spiral arms at the edges of the disk 
\citep[also seen at larger scales in][]{Gieser2024} with IRS3B having hints of spiral structure, suggesting an interaction as part of a bridge of material linking IRS3A and IRS3B \citep{Gieser2024}. 

\begin{figure*}
  \centering
\includegraphics[angle=0,width=0.99\textwidth]{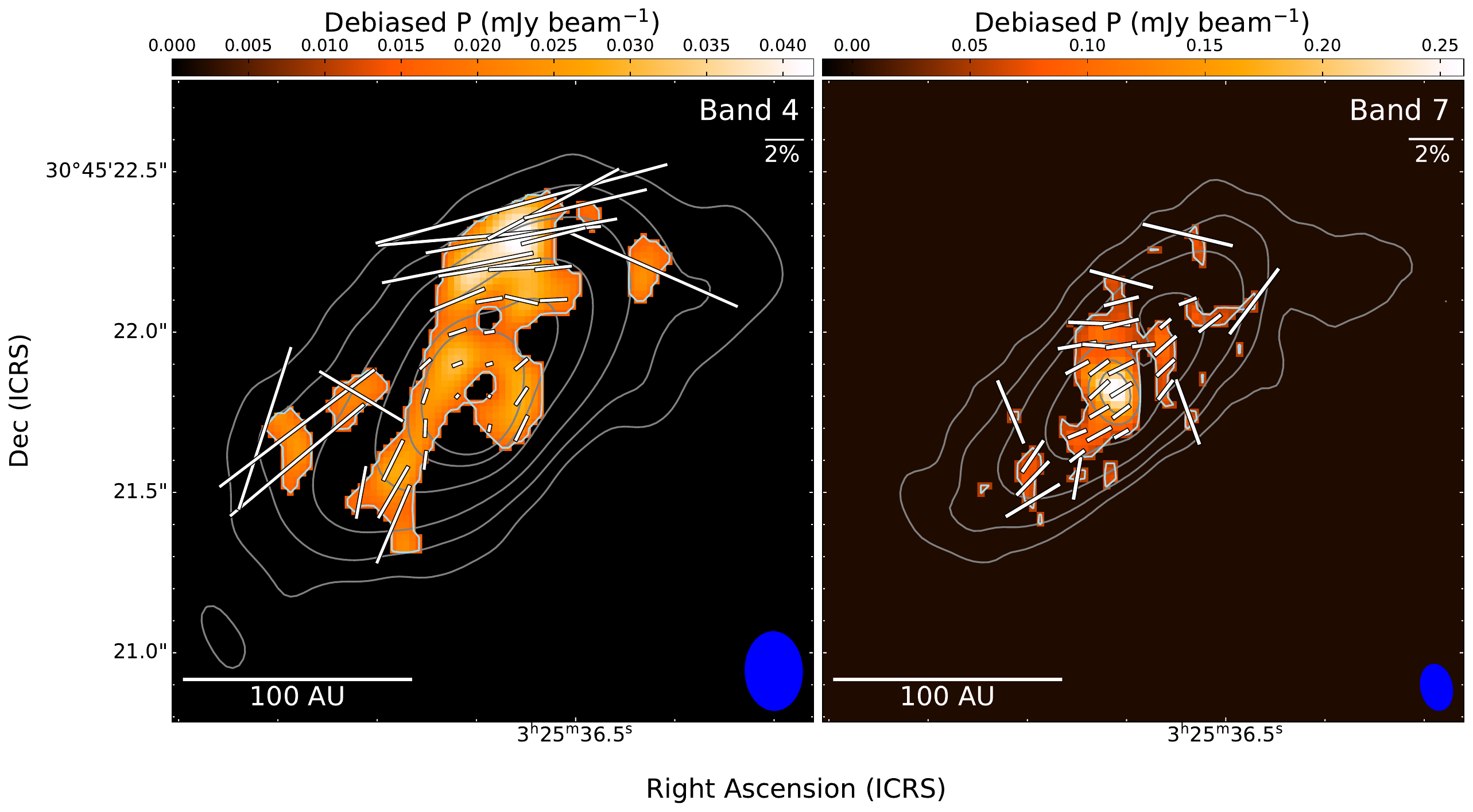}
  \caption{ALMA Band 4 and Band 7 linear polarized intensity for L1448 IRS3A. The light grey contours are Stokes I intensity (5, 10, 20, 50, 100, and 150 $\times\sigma_I$, where $\sigma_I$ is given in Table \ref{tab:maps}. The white contour is the 2$\sigma_P$ level; the linear polarized intensity is only shown when P $>$ 2$\sigma_P$ and Stokes I $>$ 10$ \sigma_I$. The line segments are the polarization angle with the length related to the percent polarization; a 2\% polarization length is given in the upper right. We plot the polarization angles with Nyquist sampling along the beam minor axis. The beams, sizes listed in Table \ref{tab:maps}, are shown in blue in the bottom right corners. 
  }
  \label{LPI_A}
\end{figure*}

Figure \ref{LPI_A} presents the polarization and polarization angle of the L1448 IRS3A dust continuum at Band 4 and Band 7. The polarization is detected in an area equal to a few beams in each case, but unlike L1448 IRS3B, the morphology in the two Bands are different. Band 4 has a polarization hole toward the center with a hint of azimuthal polarization morphology, which may be consistent with mechanical alignment, but the data are not clear.  Band 7, on the other hand, has a peak polarization at the center and a polarization morphology that is aligned with the major axis.  This is not consistent with polarization due to scattering.  If the polarization is due to magnetic field alignment, it would suggest a poloidal magnetic field that is parallel to the minor axis.

\begin{figure*}
  \centering
    \includegraphics[angle=0,width=0.50\textwidth]{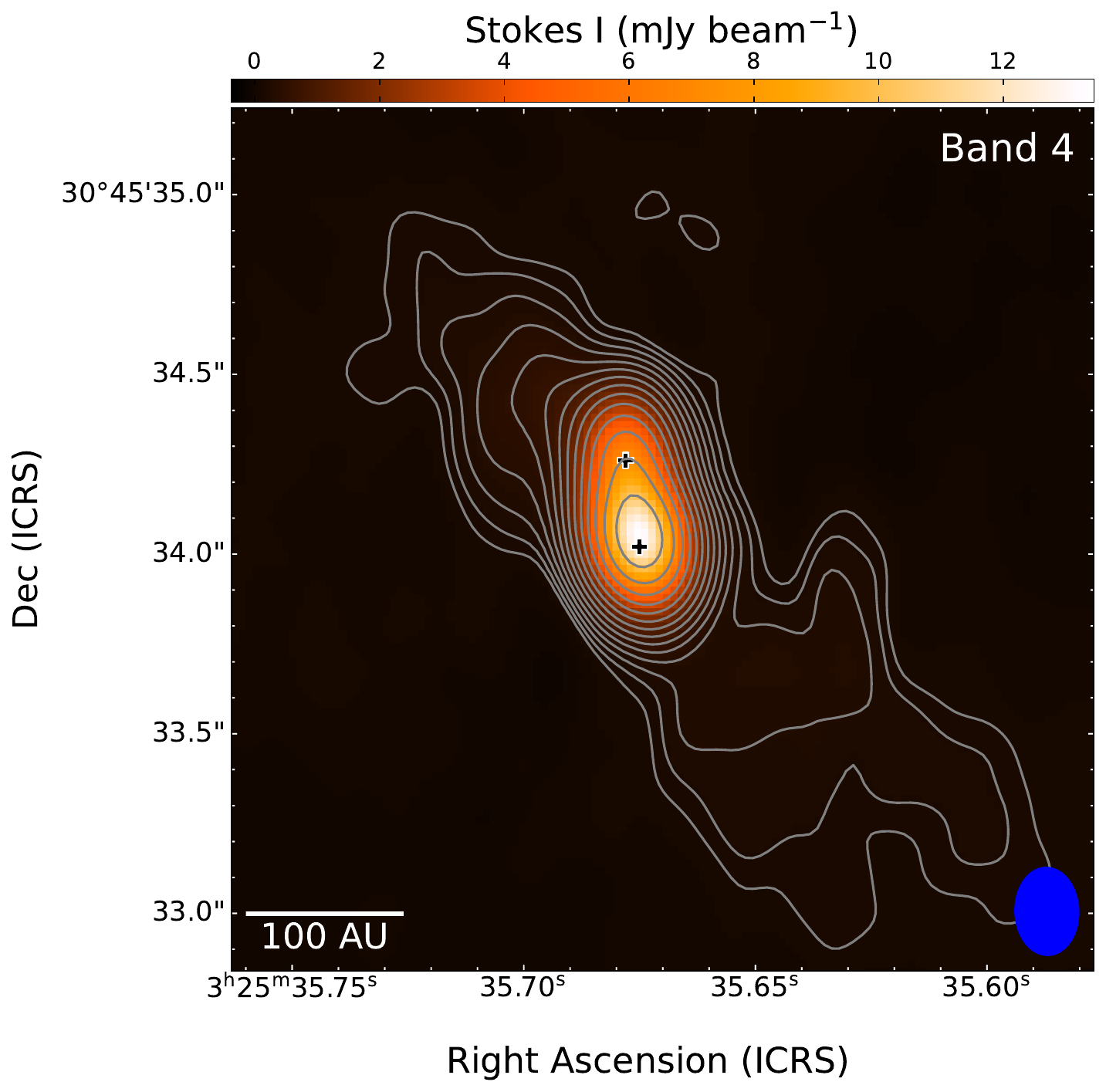}
  \caption{Band 4 continuum map of L1448 IRS3C (L1448 NW), primary beam corrected.  
  The contours are 3$\sigma\times(2)^{n/2}$, with integer n starting at 0 and using the $\sigma$ listed in Table \ref{tab:maps}, divided by 0.45, the primary beam correction factor. 
  The beam, size listed in Table \ref{tab:maps}, is shown in blue in the bottom right corner. The black + symbols mark the binary protostar locations from \cite{Reynolds2023}.
 }
  \label{stokesI_IRS3C}
\end{figure*}

Figure \ref{stokesI_IRS3C} presents the dust emission from the Class 0 source L1448 IRS3C  \citep{Looney2000,Pokhrel2023}, which is 17$\arcsec$ to the northwest from IRS3B.  The source is also often called L1448 NW \citep{Terebey1997}. L1448 IRS3C is a Class 0  close binary system with a separation of $\sim$0$\farcs$25 \citep{Reynolds2023,Pokhrel2023}. The two disks are nearly north-south.  The structure seen in Figure \ref{stokesI_IRS3C} is either a flattened envelope or a circumbinary disk embedded in a flattened filamentary structure. 

\begin{figure*}
  \centering
\includegraphics[angle=0,width=0.5\textwidth]{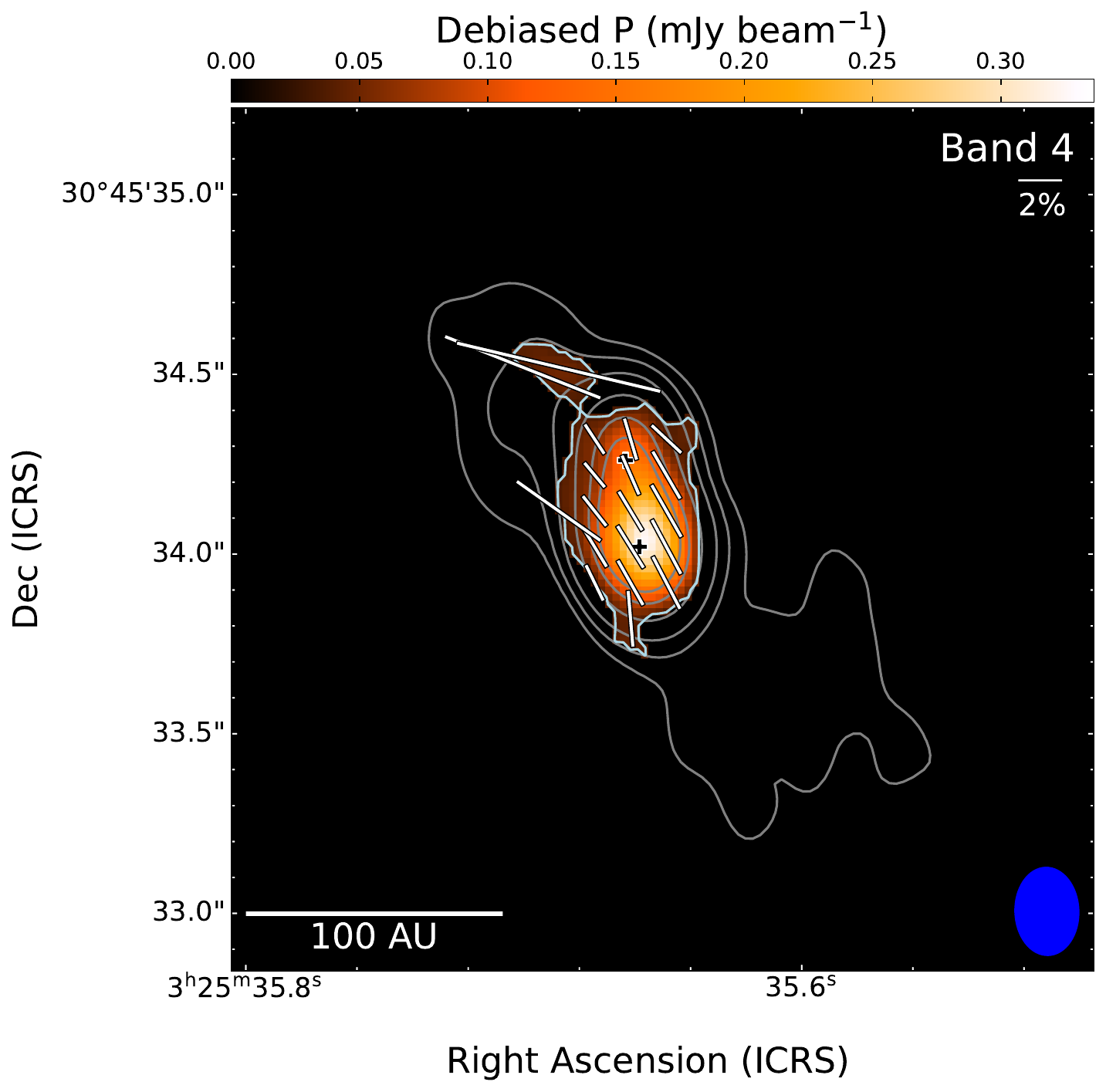}
  \caption{ALMA Band 4 linear polarized intensity toward L1448 IRS3C primary beam corrected. The light grey contours are Stokes I intensity (5, 10, 20, 50, 100, and 150 $\times\sigma_I$, where $\sigma_I$ is given in Table \ref{tab:maps}, divided by 0.45, the beam correction factor. The white contour is the 2$\sigma_P$ level; the linear polarized intensity is only shown when P $>$ 2$\sigma_P$ and Stokes I $>$ 10$ \sigma_I$. The line segments are the polarization angle with the length related to the percent polarization; a 2\% polarization length is given in the upper right. We plot the polarization angles with Nyquist sampling along the beam minor axis. 
  The beam, size listed in Table \ref{tab:maps}, is shown in blue in the bottom right corner. The black + symbols mark the binary protostar locations from \cite{Reynolds2023}.
  }
  \label{LPI_C}
\end{figure*}

Figure \ref{LPI_C} presents the polarization intensity and polarization angle of IRS3C.  The polarization is not consistent with scattering, which is common for either envelopes or circumbinary disks.  The polarization angle is aligned with the flattened envelope structure, which if due to magnetic field dust alignment, would suggest a poloidal magnetic field that is parallel to the minor axis.  However, it is important to note that the polarization detected here is outside of the ALMA 1/3 primary beam (actually just outside the 1/2 power point), making the detection less conclusive.

\begin{figure*}
  \centering
\gridline{
\leftfig{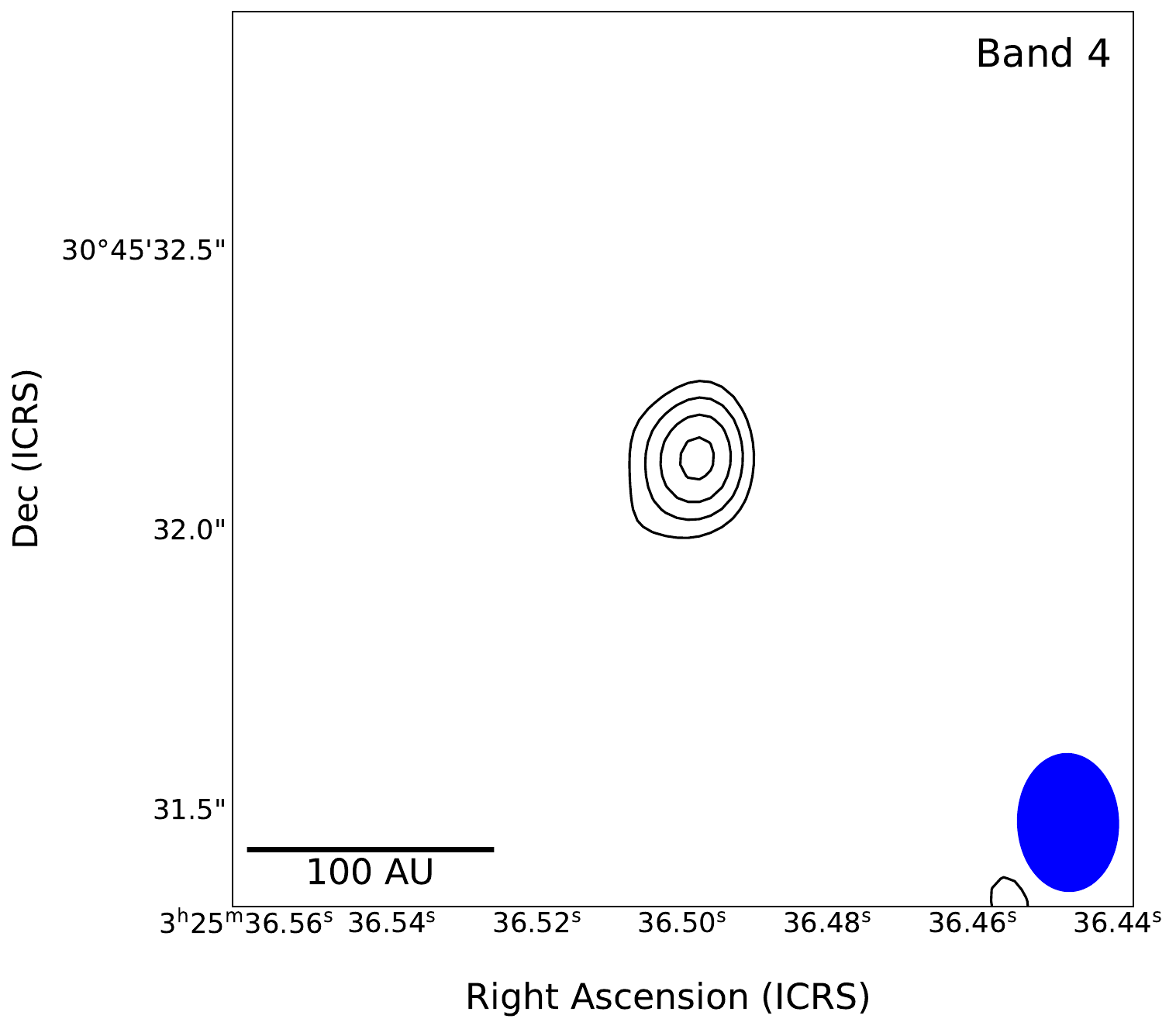}{0.48\textwidth}{}
\rightfig{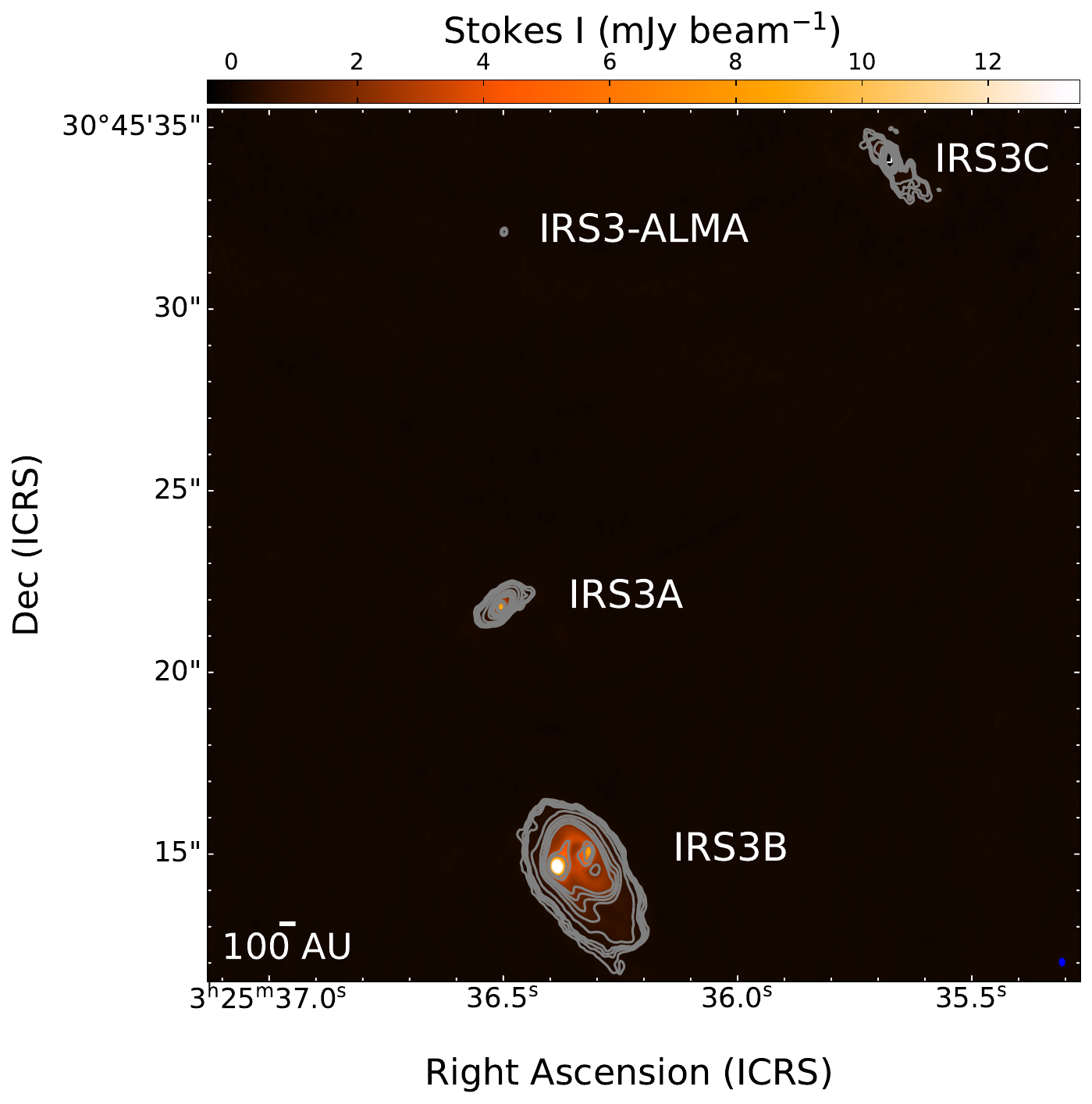}{0.52\textwidth}{}
    }
\caption{Left: Band 4 continuum map of a 6-$\sigma$ detection of a new source, labeled L1448 IRS3-ALMA, in the region, primary beam corrected.  
  The contours are $\pm$3,4, 5, and 6$\sigma$ using the $\sigma$ listed in Table \ref{tab:maps}, divided by 0.59, the beam correction factor. Note that we do not use a colormap due to the weak source brightness.
  The beam, size listed in Table \ref{tab:maps}, is shown in blue in the bottom right corner.
  Right: Zoom out of the Band 4 continuum map to show the full L1448 region with the 4 sources labeled. The grey contours are drawn using the RMS of the IRS3C map in Figure \ref{stokesI_IRS3C} $\times$ 3, 4, 5, 6, 10, 20, 30, 40, 50, 100, and 150.
 }
  \label{stokesI_newsource}
\end{figure*}

Finally, Figure \ref{stokesI_newsource} presents a 6-$\sigma$ continuum point source detection in Band 4 toward the north of the field of view (17$\farcs$5 away from IRS3B) that we label L1448 IRS3-ALMA.  L1448 IRS3-ALMA is undetected in the Perseus VANDAM survey of \cite{Tobin2016} in the VLA Ka Band ($\lambda$ = 8~mm and 1~cm) with RMS values of 0.012 and 0.01 mJy/beam, respectively.  In our Band 4 observations, the total flux is 221 $\pm$25 $\mu$Jy/beam (primary beam corrected), which means the emission is falling, consistent with dust emission.  L1448 IRS3-ALMA is also not detected in the VANDAM ALMA Band 6 observations ($\lambda$ = 1.3 mm) from \cite{Tobin2018} with an RMS of 0.33 mJy/beam (since the source is near the half-power point in their observations), which implies a spectral index, $\alpha <$ 3.2, which is also consistent with dust emission.
We could not find any source known at the L1448 IRS3-ALMA position; it is most likely a background dusty galaxy.

\section{IRS3B Polarization Rotated by 90$^\circ$}

\begin{figure*}
  \centering
\includegraphics[angle=0,width=0.70\textwidth]{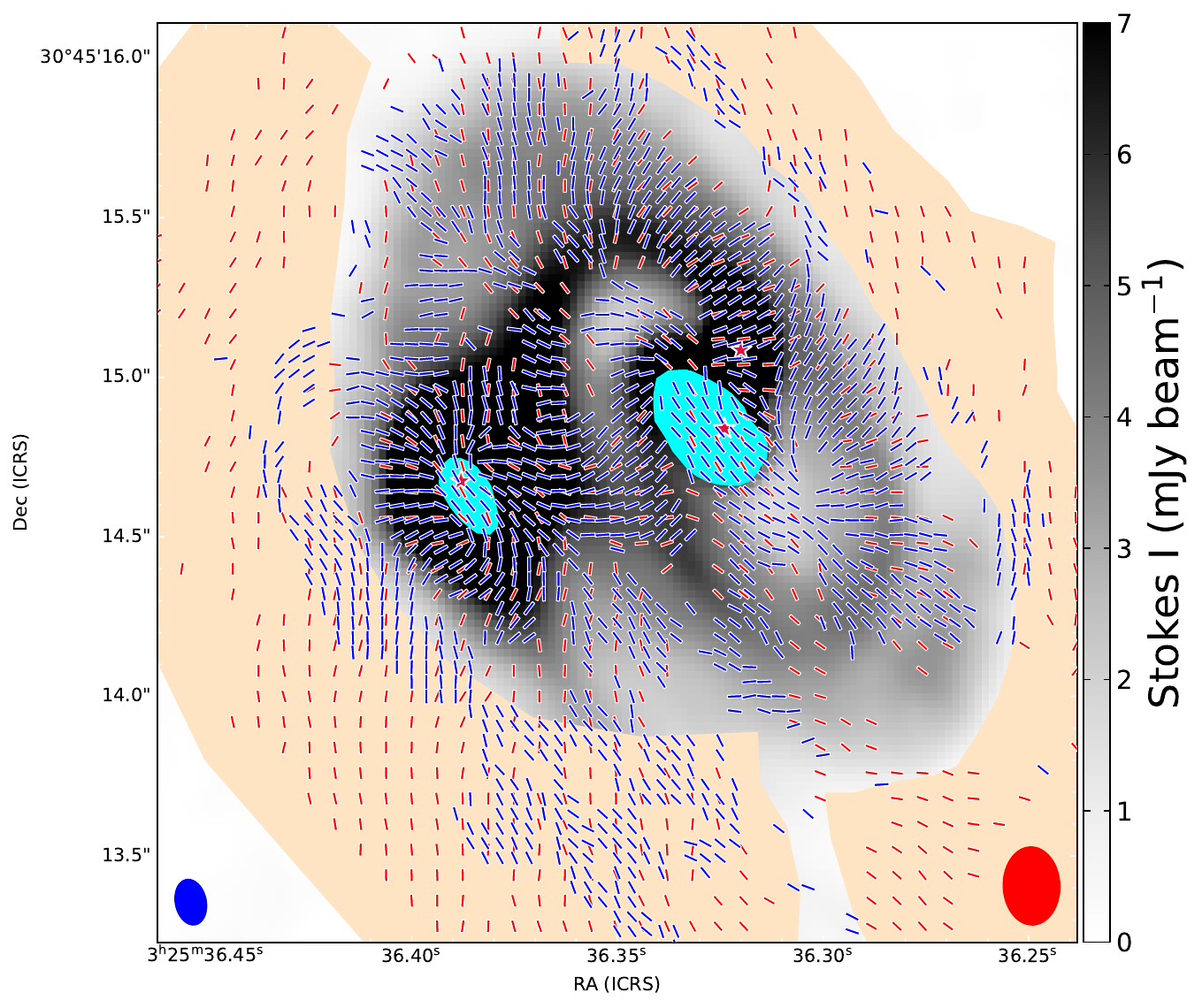}
  \caption{
  Same as Figure \ref{Regions} but with the polarization angles rotated by 90$^\circ$.
  }
  \label{RegionsRotated}
\end{figure*}

In addition to the polarization angle observations of Figure \ref{Regions}, we also show the same overlay of both Band 4 and 7, at the same resolution, with their polarization angles rotated by 90$^\circ$ in Figure \ref{RegionsRotated}. This provides a plane of the sky inferred magnetic field for  polarization due to magnetic field alignment. We posit that the tan region is likely dominated by magnetic field dust grain alignment, which would infer a magnetic field in the plane of the sky to have a more toroidal morphology.

\bibliography{bib}{}
\bibliographystyle{aasjournal}



\end{document}